\documentclass[acmsmall]{acmart}
\AtBeginDocument{%
  }
\allowdisplaybreaks
\usepackage[ruled,vlined,linesnumbered]{algorithm2e}
\usepackage{graphicx,enumitem,booktabs,amsmath}

\usepackage{wrapfig}
\usepackage{mathabx}
\usepackage{lineno}

\usepackage[para,online,flushleft]{threeparttable}
\usepackage{psfrag}
\usepackage{subfigure}
\usepackage{tikz}
\usepackage{marginnote}
\usepackage{color}
\usepackage{pdfpages}
\newcommand{\bm}[1]{\mathbf{#1}}
\usepackage{multirow}
\renewcommand{\paragraph}[1]{\smallskip\noindent{\bf {#1. }}}

\newcommand{\ed}{\textsf{ED}}
\newcommand{\heavy}{\textup{\textsf{heavy}}}
\newcommand{\light}{\textup{\textsf{light}}}

\newcommand{\Q}{\mathcal{Q}}
\newcommand{\E}{\mathcal{E}}
\newcommand{\V}{\mathcal{V}}
\newcommand{\R}{\mathcal{R}}
\newcommand{\T}{\mathcal{T}}

\renewcommand{\H}{\mathcal{H}}

\newcommand{\C}{\mathbb{C}}

\newcommand{\width}{\textsf{\upshape width}}
\newcommand{\ftd}{\textsf{\upshape FTD}}
\newcommand{\td}{\textsf{\upshape TD}~}
\newcommand{\tds}{\textsf{\upshape TD}s}
\newcommand{\subw}{\textsf{\upshape subw}}

\newcommand{\nodes}{\textsf{\upshape nodes}}
\newcommand{\OUT}{\mathrm{OUT}}
\newcommand{\freew}{\textsf{\upshape freew}}

\newcommand{\projw}{\textsf{\upshape projw}}
\newcommand{\fnfhtw}{\textsf{\upshape fn-fhtw}}
\newcommand{\fnsubw}{\textsf{\upshape fn-subw}}
\newcommand{\tOUT}{\Tilde{\OUT}}
\renewcommand{\S}{\mathcal{S}}
\renewcommand{\L}{\mathcal{L}}

\newcommand{\dom}{\mathrm{dom}}
\newcommand{\y}{\mathbf{y}}

\newcommand{\X}{\mathcal{X}}
\renewcommand{\star}{\Q_{\textup{\textsf{star}}}}

\renewcommand{\line}{\Q_{\textup{\textsf{line}}}}

\renewcommand{\matrix}{\Q_{\textup{\textsf{matrix}}}}
\renewcommand{\paragraph}[1]{\smallskip\noindent{\bf {#1. }}}

\begin{document}

    \title{Output-Optimal Algorithms for Join-Aggregate Queries}

    \author{Xiao Hu}
	\email{xiaohu@uwaterloo.ca}
	\orcid{0000-0002-7890-665X}
	\affiliation{
		\institution{University of Waterloo}
		\streetaddress{200 University Ave W}
		\city{Waterloo}
		\state{Ontario}
		\country{Canada}
		\postcode{N2L 3G1}
	} 


	\setcopyright{acmlicensed}
	\acmJournal{PACMMOD}
	\acmYear{0000} 
    \acmVolume{0000} 
    \acmNumber{0000} 
    \acmArticle{0000} \acmMonth{0000}
    \acmDOI{0000}

    \renewcommand{\shortauthors}{Xiao Hu}
    \begin{abstract}
    One of the most celebrated results of computing join-aggregate queries defined over commutative semi-rings is the classic Yannakakis algorithm proposed in 1981.
    It is known that the runtime of the Yannakakis algorithm is $O(N + \OUT)$ for any free-connex query, where $N$ is the input size of the database and $\OUT$ is the output size of the query result. This is already output-optimal. However, only an upper bound $O(N \cdot \OUT)$ on the runtime is known for the large remaining class of acyclic but non-free-connex queries. Alternatively, one can convert a non-free-connex query into a free-connex one using tree decomposition techniques and then run the Yannakakis algorithm. This approach takes $O\left(N^{\#\fnsubw} + \OUT\right)$ time, where $\#\fnsubw$ is the {\em free-connex sub-modular width} of the input query. But, none of these results is known to be output-optimal. 
    
    In this paper, we show a matching lower and upper bound $\Theta\left(N \cdot \OUT^{1- \frac{1}{\fnfhtw}} + \OUT\right)$ for computing general acyclic join-aggregate queries by {\em semiring algorithms, where $\fnfhtw$ is the free-connex fractional hypertree width} of the query. For example, $\fnfhtw=1$ for free-connex queries, $\fnfhtw =2$ for line queries (a.k.a. chain matrix multiplication), and $\fnfhtw=k$ for star queries (a.k.a. star matrix multiplication) with $k$ relations. While this measure has been defined before, we are the first to use it to characterize the output-optimal complexity of acyclic join-aggregate queries. To our knowledge, this has been the first polynomial improvement over the Yannakakis algorithm in the last 40 years and completely resolves the open question of an output-optimal algorithm for computing acyclic join-aggregate queries.
    \end{abstract}

    \maketitle

    \paragraph{Acknowledge}
    This work was supported by the Natural Sciences and Engineering Research Council of Canada -- Discovery Grant.
    
    \section{Introduction}
    \label{sec:intro}
    Join-aggregate queries defined over commutative semi-rings have wide applications in data analytical tasks.
    For example, join-aggregate queries over Boolean semiring can capture the CNF satisfiability problem, the $k$-colorability problem on graphs, the Boolean conjunctive query~\cite{abiteboul1995foundations}, the constraint satisfaction problem, and the list recovery problem in coding theory~\cite{guruswami2007algorithmic}.
    As another example, join-aggregate queries over sum-product semiring have been widely used in complex network analysis (such as clustering coefficients and transitivity ratio), discrete Fourier transforms, graph analysis (such as homomorphism and Holant problem~\cite{cai2014complexity}), counting quantified conjunctive query, and permanent computation of matrices.
    Finally, join-aggregate queries over max-product semiring can capture the maximum a posteriori problem in probabilistic graph models and maximum likelihood decoding in linear codes~\cite{aji2000generalized}. We refer interested readers to~\cite{abo2016faq, aji2000generalized, dechter1999bucket,kohlas2008semiring} for many more applications.

    Finding efficient algorithms for computing {\em join-aggregate queries} has been a holy grail in database theory since 1981. 
    Previous results have achieved two flavors of runtimes: {\em worst-case optimal}~\cite{veldhuizen2014leapfrog, ngo2012worst, khamis17:_what, yannakakis1981algorithms, atserias2008size} and {\em output-sensitive}~\cite{yannakakis1981algorithms, khamis17:_what}.
    Worst-case optimal bounds are tight only on pathological instances with huge outputs, which are rare in practice.
    In contrast, output-sensitive bounds express the runtime as a function of the input size $N$ and output size $\OUT$, which are more practically meaningful, especially for queries where the aggregation may significantly reduce the output size.
    In addition, output-sensitive algorithms can imply worst-case optimal ones: the classical Yannakakis algorithm~\cite{yannakakis1981algorithms} is the best-known example, which achieves an output-optimal bound of $O(N+\OUT)$ for {\em free-connex} queries (including {\em acyclic joins} as special 
    
    \begin{wrapfigure}{r}{0.28\linewidth}
     \vspace{-2em}
     \centering
       \includegraphics[scale=0.88]{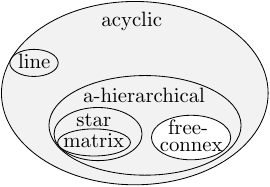}
       \vspace{-2em}
       \caption{Classification of acyclic join-aggregate queries.}
       \vspace{-1em}
     \label{fig:classification}
    \end{wrapfigure}
    
    \noindent cases).
    Although output-optimal algorithms are practically desirable, they are much more difficult to design.
    Figure~\ref{fig:classification} illustrates the relationships between different classes of acyclic join-aggregate queries.
    For the largest class of {\em acyclic but non-free-connex} queries (in shadow area in Figure~\ref{fig:classification}), 
    prior works have yet to discover an output-optimal algorithm. 
    Currently, there are only two approaches for computing these queries: (1) run the Yannakakis algorithm; (2) convert the query into a {\em free-connex one} using the {\em tree decomposition} technique and the {\em worst-case optimal join} algorithm~\cite{ngo2018worst, ngo2014skew}, and then run the Yannakakis algorithm on the tree decomposition. For (1), Yannakakis only gave an upper bound $O\left(N \cdot \OUT\right)$ on its runtime. Later, this bound has been tightened to 
    $O\left(N \cdot \OUT^{1-\frac{1}{k}}\right)$ for {\em star queries} with $k$ relations (the {\em matrix multiplication} query is the special case with two relations), which is already output-optimal~\cite{pagh14:_input}. For (2), Khamis et al. showed an upper bound $O\left(N^{\#\fnsubw} + \OUT\right)$ on its runtime, where $\#\fnsubw$ is the {\em \#free-connex submodular width} of the input query~\cite{khamis17:_what}. Both algorithms are worst-case optimal (see Appendix~\ref{appendix:yannakakis}), but neither has been shown to be {\em output-optimal}.

    In this work, we identify the free-connex fractional hypertree width ($\fnfhtw$) for join-aggregate queries. While this measure has been defined before, we are the first to use it to characterize the output-optimal complexity of acyclic join-aggregate queries. We develop a matching lower and upper bound $\Theta\left(N \cdot \OUT^{1-\frac{1}{\fnfhtw}} + \OUT\right)$ for computing general acyclic join-aggregate queries. 
    Note that $\fnfhtw=1$ for the free-connex queries and $\fnfhtw = k$ for star queries with $k$ relations, thereby generalizing the previous results on these two special cases.
    Furthermore, since this bound is output-optimal, it unifies and improves previously mentioned incomparable approaches for acyclic queries.
    As a by-product, our output-optimal algorithm for acyclic queries also yields new output-sensitive algorithms for cyclic queries, although their optimality remains unclear. In addition to our algorithmic contribution, we prove that several important notions of width identified in the literature, such as \#free-connex submodular width~\cite{khamis2020functional} and free-connex submodular width~\cite{khamis2020functional}, {\em collapse} to free-connex fractional hypertree width $\fnfhtw$ on acyclic queries.
    This surprising finding further verifies our intuition that this is the right notion to capture the output-optimal complexity of acyclic join-aggregate queries.

    \subsection{Problem Definition}
    \label{sec:problem-definition}
    \noindent {\bf Join Queries.} A (natural) {\em join} is defined as a hypergraph $q = (\V, \E)$, where the set of vertices $\V=\{x_1,\dots,x_\ell\}$ model the {\em attributes} and the set of hyperedges $\E =\{e_1,\dots, e_k\} \subseteq 2^{\V}$ model the {\em relations}. Let $\dom(x)$ be the {\em domain} of attribute $x \in \V$. Let $\dom(X) = \prod_{x \in X} \dom(x)$ be the {\em domain} of a subset $X \subseteq \V$ of attributes. An {\em instance} of $q$ is associated a set of relations $\R = \{R_e: e \in \E\}$.
    Each relation $R_e$ consists of a set of {\em tuples}, where each tuple $t \in R_e$ is an assignment that assigns a value from $\dom(x)$ to $x$ for every attribute $x\in e$. The hypergraph $q$ is called {\em self-join-free} if every relation $R_e$ is distinct. In this work, we focus on self-join-free queries. More specifically, our lower bounds assume self-join-free queries, but our algorithm can be applied to the case when self-join exists. The {\em full join result} of $q$ on $\R$, denoted as $q(\R)$, is defined as
    $\displaystyle{q(\R) = \left\{t \in \dom(\V): \forall e \in \E, \pi_e t \in R_e\right\}}$,
    i.e., all combinations of tuples, one from each relation, such that they share the same values on their common attributes. 
    
    Given a join query $q$ and an instance $\R$, the {\em effective domain} of a subset of attributes $X \subseteq \V$ is defined as the collection of tuples in $\dom(X)$ that appears in at least one full join result of $q(\R)$, i.e., the projection of $q(\R)$ onto $X$. 
     
    \paragraph{Join-Aggregate Queries} 
    A {\em join-aggregate query} is defined as a triple $\Q= (\V,\E,\y)$, where $q = (\V,\E)$ is a (natural) join, and $\y \subseteq \V$ is the set of {\em output attributes}. 
    Let $(\mathbf{D}, \oplus, \otimes, \mathbf{0}, \mathbf{1})$ be a commutative semi-ring.  We consider an instance $\R$ for $\Q$ with {\em annotated relations} \cite{green07:_proven,joglekar16:_ajar}. Every tuple $t$ is associated with an {\em annotation} $w(t) \in \mathbf{D}$. The annotation of a full join result $t \in q(\R)$ is $w(t) := \bigotimes_{e\in \E} w(\pi_e t)$. The {\em query result} of $\Q$ on $\R$ is defined as \[\Q(\R) = \bigoplus_{\V - \y} q(\R) = \left\{ (t_{\y}, w(t_{\y})) : t_\y \in \pi_\y q(\R), w(t_{\y}) = \bigoplus_{t \in q(\R): \pi_{\y} t = t_{\y}} w(t)\right\}.\]
    
    In plain language, a join-aggregate query (semantically) first computes the full join result $q(\R)$ and the annotation of each result, which is the $\otimes$-aggregate of the tuples comprising the join result.  Then it partitions $q(\R)$ into groups by the attributes in $\y$.  Finally, for each group, it computes the $\oplus$-aggregate of the annotations of the join result in that group. As mentioned, join-aggregate queries include many common database queries as special cases.  For example, if we ignore the annotations, it becomes a join-project query $\pi_{\y} q(\R)$, also known as a {\em conjunctive query}. If we take $\mathbf{D}$ be the domain of integers and set $w(t) = 1$ for every tuple $t$, it becomes the {\tt COUNT(*) GROUP BY} $\y$ query; in particular, if $\y = \emptyset$, the query computes the full join size $|q(\R)|$. If we take $\V = \{A,B,C\}$ with $\y = \{A,C\}$, and $\E= \{\{A,B\}, \{B,C\}\}$, it becomes the {\em matrix multiplication query}. If we take $\V = \{A_1,A_2,\cdots, A_k, B\}$ with $\y = \{A_1,A_2,\cdots, A_k\}$, and $\E= \{\{A_1,B\}, \cdots \{A_k,B\}\}$, it becomes a {\em star query}. If we take $\V = \{A_1,A_2,\cdots,A_{k+1}\}$ for $k \ge 3$ with $\y = \{A_1,A_{k+1}\}$, and $\E= \{\{A_1,A_2\}, \{A_2,A_3\}, \cdots, \{A_k,A_{k+1}\}\}$, it becomes a {\em line query}.
    
    Below, if not specified, a query always refers to a join-aggregate query. We use $N = \sum_{e \in \E} |R_e|$ to denote the {\em input size} of $\R$ and $\OUT = |\Q(\R)|$ to denote the {\em output size} of $\Q$ over $\R$. We study the data complexity of this problem by assuming the query size (i.e., $|\V|$ and $|\E|$) as constants. 

    \paragraph{Model of Computation} We use the standard RAM model with uniform cost measures. A tuple or a semiring element is stored in a word. Copying one semiring element or combining two semiring elements via a semiring operation ($\oplus$ and $\otimes$) can be done in $O(1)$ time. Inheriting from \cite{pagh14}, we confine ourselves to {\em semiring algorithms} that work with semiring elements as an abstract type and can only copy them from existing semiring elements or combine them using $\oplus$ or $\otimes$. No other operations on semi-ring elements are allowed, such as division, subtraction, or equality check.

    \paragraph{Output-Optimality} To establish {\em output-optimality} for algorithm design, we consider a unified output-sensitive upper and lower bound in Definition~\ref{def:bound}.
    \begin{definition}[Output-sensitive Bound]
        \label{def:bound}
        For a self-join-free query $\Q$, let $f(\Q)$ be the smallest exponent such that for any parameters $1 \le N$ and $\displaystyle{\OUT \le \max_{\R' \in \Re(N)}\left|\Q(\R')\right|}$, a semi-ring algorithm exists that can compute $\Q(\R)$ for any instance $\R$ of input size $N$ and output size $\OUT$ within $O\left(N \cdot \OUT^{1-\frac{1}{f(\Q)}} + \OUT \right)$ time, where $\Re(N)$ is the set of all instances over $\Q$ of input size $N$.
    \end{definition}

    This bound is a monotonic function of $f(\Q)$. A smaller $f(\Q)$ implies a smaller runtime and, therefore, a better upper bound. For example, $f(\Q) \le 1$ implies the upper bound $O\left(N + \OUT \right)$; and $f(\Q) \le +\infty$ implies the upper bound $O\left(N \cdot \OUT\right)$.
    This definition can also express lower bounds. For example, $f(\Q) \ge 1$ implies the lower bound $\Omega\left(N + \OUT \right)$.
    In this work, we characterize exactly $f(\cdot)$ for acyclic queries with both lower and upper bounds.

    \subsection{Our New Lower Bound for Acyclic Queries}
    \label{sec:lb}
    Prior work has provided lower bounds of $f(\Q)$. First, $f(\Q) \ge 1$ for all queries since any algorithm must read the input data and output all query results. 
    Pagh et al. showed $f(\Q) \ge k$ for star queries with $k$ relations \cite{pagh14}. Hu identified the {\em free-width} for an acyclic query $\Q$, denoted as $\freew(\Q)$, and showed that $f(\Q) \ge \freew(\Q)$~\cite{hu2024fast}. In this paper, we prove: 
     \begin{theorem}
     \label{the:lb}
         For any acyclic query $\Q$, $f(\Q) \ge \fnfhtw(\Q)$. 
     \end{theorem}
    \noindent
    We next give some simple observations to understand our significant improvement over~\cite{hu2024fast} and defer the detailed comparison between $\fnfhtw(\Q)$ and $\freew(\Q)$ to Section~\ref{sec:fnfhtw}. First, $\fnfhtw(\Q) \ge \freew(\Q) \ge 1$ for all queries. In some cases, $\fnfhtw(\Q) = \freew(\Q)$, such as free-connex queries, line queries, and star queries.
    But, for many other cases, $\fnfhtw(\Q) > \freew(\Q)$. 

    This $\fnfhtw$-dependent lower bound can be broken beyond semiring algorithms. For example, some works use fast matrix multiplication techniques to speed up conjunctive queries (as a special case of join-aggregate queries defined over Boolean semiring) processing~\cite{amossen2009faster, deep2020fast, abboud2024time, hu2024fast} or graph pattern search (as a special case of self-joins)~\cite{bjorklund2014listing, jin2024listing, dalirrooyfard2023listing}. However, we cannot apply these techniques to arbitrary join-aggregate queries, since a general semiring does not necessarily have an additive inverse (such as the tropical semi-ring), 
    so we won't pursue this dimension further in this paper.

    \subsection{Our New Upper Bound for Acyclic Queries}
    \label{sec:up}
    In this paper, we propose a new algorithm by exploring a hybrid version of the Yannakakis algorithm for computing acyclic queries, and therefore we can prove: 
    \begin{theorem}
    \label{the:up}
        For any acyclic query $\Q$, $f(\Q) \le \fnfhtw(\Q)$.
    \end{theorem}
    \noindent
    Combine Theorem~\ref{the:lb} and Theorem~\ref{the:up}, we obtain a full understanding of $f(\cdot)$ for acyclic queries:
    \begin{corollary}
        \label{cor:optimal}
        For any acyclic query $\Q$, $f(\Q) = \fnfhtw(\Q)$.
    \end{corollary}

    \noindent {\bf Comparison with~\cite{yannakakis1981algorithms}.} One question remains for the Yannakakis algorithm:
    \textit{Is the unsatisfactory upper bound $O(N \cdot \OUT)$ due to a fundamental limitation of the algorithm itself or just because we do not have a tight analysis of its runtime?} The runtime bound of the Yannakakis algorithm has been tightened on free-connex queries and star queries~\cite{pagh14:_input}. 
    Recently, Hu showed that the Yannakakis algorithm indeed requires $\Theta(N \cdot \OUT)$ time for {\em line queries}~\cite{hu2024fast}, 
    which has first demonstrated the limitation of the Yannakakis algorithm.
    We provide a tight analysis for all acyclic queries. More specifically, the runtime bound can be tightened to $O\left(N \cdot \OUT^{1-\frac{1}{\fnfhtw(\Q)}} + \OUT\right)$ for {\em a-hierarchical queries}. In contrast, $O(N \cdot \OUT)$ is tight for all non-a-hierarchical queries.  As shown in Figure~\ref{fig:classification}, a-hierarchical queries include free-connex and star queries, but not line queries. Hence, our new algorithm strictly outperforms the Yannakakis algorithm on all non-a-hierarchical queries.

    \paragraph{Comparison with~\cite{khamis17:_what}}
    Another approach (even applying for cyclic queries) that converts a query into a free-connex one and then runs the Yannakakis algorithm takes $O\left(N^{\#\fnsubw(\Q)} + \OUT\right)$ time, where $\#\fnsubw(\Q)$ is the \#free-connex sub-modular width of the input query $\Q$~\cite{khamis17:_what}. 
    As shown in Lemma~\ref{lem:equivalence}, both notions of width surprisingly collapse on all acyclic queries!    
    In Appendix~\ref{appendix:yannakakis}, we prove this algorithm is no better (and in many cases polynomially worse) than our new algorithm. 

    \paragraph{Comparison with \cite{deep2024output}} In an independent work from ours, Deep et al. identified the {\em project-width} for an acyclic query $\Q$, denoted as $\projw(\Q)$, and showed that $f(\Q)\le \projw(\Q)$ \cite{deep2024output}. 
    Again, we give some observations to understand our advantages over this upper bound and defer a detailed comparison between $\fnfhtw(\Q)$ and $\projw(\Q)$ to Section~\ref{sec:fnfhtw}. First, $\fnfhtw(\Q)\le \projw(\Q)$ for all queries. For a-hierarchical queries, $\fnfhtw(\Q) = \projw(\Q)$. For line queries with $k$ relations, $\fnfhtw(\Q) = 2 < \projw(\Q) = k$.
    Also, for many other cases, $\fnfhtw(\Q) < \projw(\Q)$. 

    \smallskip
    We summarize the runtime of our new algorithm and comparable algorithms in Figure~\ref{fig:summary}.
    
    \subsection{Implications to Cyclic Queries}
    \label{sec:implications}

    The common approach for tackling cyclic queries is to convert them into acyclic ones using tree decomposition techniques and the worst-case optimal join algorithm, and then run the Yannakakis algorithm on the tree composition. This approach takes $O\left(N^{\#\subw(\Q)} \cdot \OUT \right)$ time, where $\#\subw(\Q)$ is the \#sub-modular width of the input query $\Q$~\cite{marx2013tractable}. 
    If restricting tree decompositions to be free-connex, this takes $O\left(N^{\#\fnsubw(\Q)} + \OUT\right)$ time. Note that $\#\subw(\Q) \le \#\fnsubw(\Q)$ for all queries. These two results are incomparable unless the value of $\OUT$ is known. By replacing the Yannakakis algorithm with our new output-optimal algorithm for acyclic queries, we can get new output-sensitive algorithms for cyclic queries. However, their optimality is unclear due to our limited understanding of the lower bounds, which we leave as an open question.
    \begin{figure*}[t]  
        \begin{tabular}{c|c|c|c}
        \toprule
        Join-Aggregate & \multirow{2}{*}{Yannakakis~\cite{yannakakis1981algorithms, khamis17:_what}} & \multirow{2}{*}{Deep et al.~\cite{deep2024output}} & \multirow{2}{*}{\color{red}Our Algorithm}\\
        Query &  &  & \\
        \hline
        \multirow{2}{*}{a-hierarchical} & \multicolumn{3}{c}{\multirow{2}{*}{\color{red}$N \cdot \OUT^{1-\frac{1}{\#\fnsubw}}$}}\\
        & 
        \multicolumn{2}{c}{} \\
        \hline
        \multirow{2}{*}{Line} & \multirow{2}{*}{$\min\left\{N \cdot \OUT, N^2\right\}$} & \multirow{2}{*}{$N \cdot \OUT^{1-\frac{1}{k}}$} & \multirow{2}{*}{$N \cdot \sqrt{\OUT}$} \\
        &  & & \\
        \cline{1-4}
        \multirow{3}{*}{Acyclic} & \multirow{3}{*}{$\min\begin{cases}&\hspace{-1em} N \cdot \OUT \\
        &\hspace{-1em} N^{\#\fnsubw} + \OUT
        \end{cases}$} & {\multirow{3}{*}{$N \cdot \OUT^{1-\frac{1}{\projw}} + \OUT$}}  & {\color{red} \multirow{3}{*}{$N \cdot \OUT^{1-\frac{1}{\fnfhtw}} + \OUT$}} \\
        & & & \\
        & & & \\
        \bottomrule
    \end{tabular}
    \vspace{-1em}
    \caption{Comparison between previous and our new upper bounds. All results are in $\Theta(\cdot)$. $N$ is the input size, and $\OUT$ is the output size. $k$ is the number of relations. 
    $\fnfhtw$ is the free-connex fractional hypertree width (Definition~\ref{def:out-width}). $\projw$ is the project-width (Definition~\ref{def:free-width}). $\#\fnsubw$ is the \#free-connex submodular width. As shown in Lemma~\ref{lem:width-comparison-1}, $\#\fnsubw(\Q) = \fnfhtw(\Q)$ for any acyclic query $\Q$.} 
    \label{fig:summary}
    \end{figure*}

    \subsection{Organization of This Paper} 
    \label{sec:organization}
    Our paper is organized as follows. In Section~\ref{sec:preliminary}, we introduce the preliminaries. In Section~\ref{sec:fnfhtw}, we define the free-connex fractional hypertree width and investigate its properties for acyclic queries. In Section~\ref{sec:warmup}, we review the Yannakakis algorithm and introduce our algorithm for line queries as an introductory example. Finally, we present our algorithm for general acyclic queries in Section~\ref{sec:acyclic}. 

    \section{Preliminaries}
    \label{sec:preliminary}
  
    \subsection{Fractional Edge Covering and AGM Bound}
    \label{sec:basic}
    
    For a join query $q = (\V,\E)$, we use $\E_x = \{e \in \E: x \in e\}$ to denote the set of relations containing attribute $x$. An attribute $x \in \V$ is {\em unique} if $|\E_x|=1$, and {\em joint} otherwise. For a subset of attributes $S \subseteq \V$, we use $q[S] = (S, \E[S])$ to denote the sub-query induced by $S$, where $\E[S] = \{e \cap S: e\in \E\}$. A {\em fractional edge covering} is a function $\rho: \E \to [0,1]$ such that $\sum_{e: A \in e} \rho(e) \ge 1$ for each attribute $A \in \V$. The {\em fractional edge covering number} of $q$, denoted as $\rho^*(q)$, is defined as the minimum sum of weight over all possible fractional edge coverings $\rho$ for $q$, i.e., $\rho^*(q) = \min_{\rho}\sum_{e \in \E} \rho(e)$. For a join query $q = (\V,\E)$ and any parameter $N \in \mathbb{Z}^+$,
    the AGM bound~\cite{atserias2008size} states that the maximum number of join results produced by any instance of input size $N$ is $\Theta(N^{\rho^*(q)})$.  For a join-aggregate query $\Q = (\V,\E,\y)$ with $q = (\V,\E)$ and any parameter $N \in \mathbb{Z}^+$, the maximum number of join results produced by any instance of input size $N$ is $\Theta(N^{\rho^*(q[\y])})$

     \subsection{Tree Decompositions}
     \label{sec:td}
        A {\em tree decomposition} of a query $\Q = (\V, \E, \y)$ is a pair $(\T, \chi)$, where $\mathcal{T}$ is a tree and $\chi: \textsf{nodes}(\T) \to 2^\V$ is a mapping
        from the nodes of $\mathcal{T}$ to subsets of $\V$, that satisfies the following properties:
        \begin{itemize}[leftmargin=*]
            \item For each relation $e \in \E$, there is a node $u \in \nodes(\mathcal{T})$ such that $e \subseteq \chi(u)$.
            \item For each attribute $x \in \V$, the set $\{u \in \nodes(\mathcal{T}): x \in \chi(u)\}$ forms a connected sub-tree of $\T$.
        \end{itemize}
        
        Each set $\chi(u)$ is called a {\em bag} of the \textsf{TD}. Wlog, we assume $\chi(u) \neq \chi(u')$ for any pair of nodes $u,u' \in \nodes(\T)$. 
        The {\em width} of $(\mathcal{T}, \chi)$ noted as $\width(\mathcal{T}, \chi)$ is defined as
        \begin{equation*}
            \label{eq:width-td}
            \width(\mathcal{T}, \chi) \quad = \quad \max_{u \in \nodes(\T)} \quad \rho^*\left(q[\chi(u)]\right)
        \end{equation*}
        i.e., the maximum fractional edge covering number of the sub-queries induced by all bags in $\mathcal{T}$.
        A \td $(\T, \chi)$ is {\em free-connex} if there is a connected subtree $S$ of $\T$ such that $\bigcup_{u \in \nodes(S)} \chi(u) = \y$, i.e., the union of attributes appearing in $S$ is exactly the output attributes. $S$ is called a {\em connex} of $\T$.    
      
    \subsection{Classification of Queries}
    \label{sec:classification}
    \noindent {\bf Acyclic~\cite{beeri1983desirability,idris17:_dynam}.} A query is acyclic if and only if it has a width-1 \textsf{TD}.
    
    \begin{wrapfigure}{r}{0.26\linewidth}
      \vspace{-3em}
        \centering
     \includegraphics[scale=1.0]{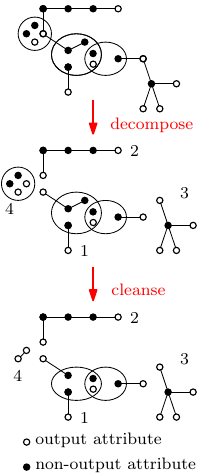}
        \caption{An illustration of the decompose and cleanse process. Hollow or solid dots are attributes. Lines or circles are relations. }
        \label{fig:pre-processing}
        \vspace{-3em}
      \end{wrapfigure}
    \paragraph{Free-connex~\cite{bagan2007acyclic}} A query is free-connex if and only if it has a width-1 free-connex \textsf{TD}.

    \paragraph{$\exists$-Connected} The {\em existential connectivity} of a query $\Q = (\V,\E,\y)$ is defined on a graph $G^\exists_\Q$, where each relation $e \in \E$ is a vertex, and an edge exists between $e, e' \in \E$ if $e\cap e'- \y \neq \emptyset$. $\Q$ is {\em $\exists$-connected} if $G^\exists_\Q$ is connected, and {\em $\exists$-disconnected} otherwise. If $\Q$ is $\exists$-disconnected, we can {\em decompose} it as follows. We find all connected components of $G^\exists_\Q$, in which the set of relations corresponding to the set of vertices in one connected component of $G^\exists_\Q$ form a connected subquery of $\Q$. There are four $\exists$-connected sub-queries of $\Q$: $1$ is a complicated acyclic query, $2$ is a line query, $3$ is a star query, and $4$ is a single relation. 

    \paragraph{Cleansed}A query $\Q$ is {\em cleansed} if every unique attribute is an output attribute, and there exist no relations whose attributes are fully contained by another one. If $\Q$ is not cleansed, we can {\em cleanse} it by iteratively removing a unique non-output attribute or a relation whose attributes are fully contained by another. The resulting query is the {\em cleansed} version of $\Q$. In Figure~\ref{fig:pre-processing}, attributes and relations in blue are removed by the cleanse process.
     
     \smallskip \noindent {\bf Separated.} A query $\Q=(\V,\E,\y)$ is {\em separated} if every output attribute is unique, every unique attribute is an output attribute and for each $e \in \E$ with $e \cap \y \neq \emptyset$, there exists $e' \in \E - \{e\}$ with $e - \y \subseteq e'$. Among these $4$ queries, all of $2, 3, 4$ are separated, but $1$ is not.

    \section{Free-connex Fractional Hypertree Width}
    \label{sec:fnfhtw}
    In this section, we give a structural definition of free-connex fractional hypertree width for general queries. For acyclic queries, we characterize an equivalent procedural definition, which is fundamental in facilitating comparisons between free-connex fractional hypertree width with other notions of width in Section~\ref{sec:comparison}, and inspiring our output-optimal algorithm in Section~\ref{sec:acyclic}.

    \subsection{Definition}
    \label{sec:def-fnfhtw}
    The structural definition of free-connex fractional hypertree width is defined on free-connex tree decompositions. 
    Let $\ftd(\Q)$ denote the set of all free-connex \tds~for a query $\Q$.

    \begin{definition}[Free-connex Fractional Hypertree Width (\fnfhtw)]
    \label{def:out-width}
        For any query $\Q = (\V,\E,\y)$, its free-connex fractional hypertree width $\fnfhtw(\Q)$ is defined as:
        \begin{equation}
        \label{eq:out-width}
            \fnfhtw(\Q) \quad = \quad \min_{(\T,\chi) \in \ftd(\Q)} \quad \width(\T,\chi)
        \end{equation}
        i.e., the minimum width of all possible free-connex \tds.
    \end{definition}

    It is easy to see $\fnfhtw(\Q) \ge 1$ for all queries; and $\fnfhtw(\Q) = 1$ for free-connex queries. We show that $\fnfhtw$ is preserved by the cleanse process in Lemma~\ref{lem:out-width-cleansed}. When restricting our scope to acyclic queries, we prove two important properties of \fnfhtw~in Lemma~\ref{lem:fn-fhtw-disconnected} and Lemma~\ref{lem:fn-fhtw-connected}, which serves as a procedural definition of \fnfhtw. Their proofs are rather technical and deferred to Appendix~\ref{appendix:fnfhtw}.   

    \begin{lemma}
    \label{lem:out-width-cleansed}
        For any query $\Q$, $\fnfhtw(\Q)= \fnfhtw(\Q')$, where $\Q'$ is cleansed version of $\Q$.
    \end{lemma}

    \begin{lemma}
        \label{lem:fn-fhtw-disconnected}
        For any acyclic query $\Q = (\V,\E,\y)$, if $\Q$ is $\exists$-disconnected with $\exists$-connected subqueries $\Q_1,\Q_2,\cdots, \Q_h$, $\fnfhtw(\Q)= \max_{i \in [h]} \fnfhtw(\Q_i)$.
    \end{lemma}

    \begin{lemma}
        \label{lem:fn-fhtw-connected}
        For any acyclic query $\Q = (\V,\E,\y)$, if $\Q$ is $\exists$-connected, $\fnfhtw(\Q)= \rho^*\left(q[\y]\right)$, i.e., the fractional edge covering number of the sub-query induced by output attributes, for $q = (\V,\E)$. 
    \end{lemma}

    \begin{corollary}
        \label{cor:recursive}
        For any acyclic query $\Q = (\V,\E,\y)$, its free-connex fractional hypertree width $\fnfhtw(\Q)$ is equivalently defined as:
        \begin{itemize}[leftmargin=*]
            \item If $\Q$ is $\exists$-disconnected with $\exists$-connected subqueries $\Q_1,\Q_2,\cdots, \Q_h$, $\fnfhtw(\Q)= \max_{i \in [h]} \fnfhtw(\Q_i)$. 
            \item If $\Q$ is $\exists$-connected, $\fnfhtw(\Q)= \rho^*\left(q[\y]\right)$, i.e., the fractional edge covering number of the sub-query induced by output attributes, for $q = (\V,\E)$. 
        \end{itemize}
    \end{corollary}

    \subsection{Comparison with Other Notions of Width}
    \label{sec:comparison}
    \noindent {\bf Free-width and Project-width.} We first review the definitions for {\em free-width} and {\em project-width}:
    \begin{definition}[Free-width~\cite{hu2024fast} and Project-width~\cite{deep2024output}]
    \label{def:free-width}
        For any acyclic query $\Q = (\V,\E,\y)$, its free-width $\freew(\Q)$ and project-width $\projw(\Q)$ are defined as follows:
        \begin{itemize}[leftmargin=*]
        \item If $\Q$ is $\exists$-disconnected with $\exists$-connected subqueries $\Q_1,\Q_2,\cdots, \Q_h$, $\freew(\Q)= \max_{i \in [h]} \freew(\Q_i)$, and $\projw(\Q)= \max_{i \in [h]} \projw(\Q_i)$.
        \item If $\Q$ is $\exists$-connected but not cleansed, $\freew(\Q)= \freew(\Q')$ and $\projw(\Q)= \projw(\Q')$, where $\Q'$ is the cleansed version of $\Q$.
        \item If $\Q$ is $\exists$-connected and cleansed, $\freew(\Q) = \displaystyle{\left|\{e\in \E:e \cap \V_\bullet \neq \emptyset\}\right|}$,
        where $\V_\bullet$ is the set of unique (output) attributes in $\Q$; and $\projw(\Q) = |\E|$. 
    \end{itemize}
    \end{definition}

    \begin{lemma}
        \label{lem:width-comparison-1}
        For any acyclic query $\Q$, $\freew(\Q) \le \fnfhtw(\Q) \le \projw(\Q)$.
    \end{lemma}

    These three notions of width share the exact definition if $\Q$ is $\exists$-disconnected or not cleansed. The only difference comes when $\Q$ is $\exists$-connected and cleansed. 
    In this case, it is easier to see that $\freew(\Q) \le \fnfhtw(\Q) \le \projw(\Q)$. First, every relation containing a unique attribute (which must be an output attribute because $\Q$ is cleansed) should be assigned a weight $1$ in any fractional edge covering of $q[\y]$. This is why $\freew(\Q) \le \fnfhtw(\Q)$ holds for all queries. Moreover, assigning all relations with weight $1$ forms a trivial fractional edge covering for $q[\y]$, hence $\fnfhtw(\Q) \le |\E| = \projw(\Q)$. In Example~\ref{exp:width}, we show a query $\Q$ with $\freew(\Q) < \fnfhtw(\Q) < \projw(\Q)$. 

    \begin{wrapfigure}{r}{0.23\linewidth}
       \centering
       \vspace{-0.5em}
        \includegraphics[scale=1.3]{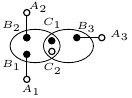}
        \vspace{-1em}
        \caption{An example query $\Q$ with $\freew(\Q)<$ $\fnfhtw(\Q)$ $<$ $\projw(\Q)$.}
        \label{fig:Q3}
        \vspace{-2em}
    \end{wrapfigure}

    \begin{example}
    \label{exp:width} \noindent Consider the query $\Q = (\V,\E,\y)$ in Figure~\ref{fig:Q3}, where $\V = \{A_1,A_2,A_3,B_1,B_2,B_3,C_1,\\C_2\}, \E = \left\{e_1=\{A_1,B_1\}, e_2=\{A_2,B_2\},e_3 =\{A_3,B_3\}, e_4=\{B_1,B_2,C_1,C_2\}, e_5=\{B_3,C_1,C_2\}\right\}$, and $\y =\{A_1, A_2, A_3, C_2\}$. Note that $A_1, A_2, A_3$ are unique output attributes. All of $e_1,e_2,e_3$ contain some unique output attribute(s), so $\freew(\Q) = 3$. The sub-query has a free-connex $\td$ with width $4$ (which is also minimum), so $\fnfhtw(\Q) = 4$. This query has $5$ relations, so $\projw(\Q)=5$.
    \end{example}

    \noindent {\bf \#Free-connex submodular width.} The \#free-connex submodular width ($\#\fnsubw$) is also defined based on tree decompositions but in a rather complicated formula. We review its formal definition in Appendix~\ref{appendix:fnfhtw}. First, we can show $\#\fnsubw(\Q)\le \fnfhtw(\Q)$ for all queries. From~\cite{khamis17:_what}, we can further draw a clear ordering by additionally involving submodular width ($\textsf{subw}$), free-connex submodular width ($\textsf{fn-subw}$), and \#submodular width ($\#\textsf{subw}$) as follows:
    \begin{lemma}
    \label{lem:general-ordering}
         For any query $\Q$, $\subw(\Q)\le\{\#\subw(\Q),\fnsubw(\Q)\}\le\#\fnsubw(\Q)\le\fnfhtw(\Q)$.
    \end{lemma}
    Surprisingly, some widths collapse when we restrict our scope to acyclic queries!
     \begin{lemma}
    \label{lem:equivalence}
        For any acyclic query $\Q$, $\fnsubw(\Q) = \#\fnsubw(\Q) = \fnfhtw(\Q)$.
    \end{lemma}

    The proof of Lemma~\ref{lem:equivalence} is also technical and deferred to Appendix~\ref{appendix:fnfhtw}.  
    This result also illustrates the fundamental difference between existing algorithms and our new algorithm. On acyclic queries, the algorithm in \cite{khamis17:_what} picks one free-connex \td and performs computation according to it, and in contrast, our algorithm picks a set of free-connex \tds~ and applies them to different sub-instances. The power of this hybrid strategy will become much clearer in our next section.

    \section{Warm Up: Yannakakis Revisited and Line Query}
    \label{sec:warmup}
   In this section, we first review the Yannakakis algorithm. As the simplest query on which the Yannakakis algorithm is not optimal, we next show our new algorithm for line queries (a.k.a. chain matrix multiplication) that have separately received a lot of attention~\cite{godbole1973efficient, nishida2011accelerating, barthels2018generalized, lin2024efficient}:
     $$\line = \bigoplus_{A_2,A_3,\cdots, A_{k}} R_1(A_1,A_2)\Join R_2(A_2,A_3) \Join \cdots \Join R_k(A_k,A_{k+1})$$
    aiming to illustrate some high-level ideas behind our general algorithm in Section~\ref{sec:acyclic}.

    \subsection{Yannakakis 
    Algorithm Revisited}
    \label{sec:Yannakakis}
    
    Suppose we are given an acyclic query $\Q = (\V,\E,\y)$ with $q = (\V,\E)$ and an instance $\R$ for $\Q$. Wlog, assume there exists no pair of relations $e,e'\in \E$ such that $e\subseteq e'$; otherwise, we simply update $R_{e'}$ with $R_{e'} \Join R_{e}$ and remove $e$ from $\E$. Let $(\T,\X)$ be a width-1 \td\ for $\Q$. For each node $u \in \nodes(\T)$, we define a derived relation $R_u$ as follows: if there exists some $e \in \E$ with $\chi(u)=e$, $R_u:= R_e$; otherwise, $R_u :=\Join_{e \in \E} \pi_{e \cap \chi(u)} R_e$ in which each tuple $t \in R_u$ has its annotation $w(t)=1$. 
    As $\rho^*(q[\chi(u)])=1$, there must exist some relation $e \in \E$ with $\chi(u) \subseteq e$. Hence, $R_u$ contains $O(N)$ tuples and can be computed in $O(N)$ time. The Yannakakis algorithm consists of two phases (with its pseudocode given at Algorithm~\ref{alg:yannakakis} in Appendix~\ref{appendix:yannakakis}):

    \begin{figure}[t]
         \centering
         \includegraphics[scale=1.0]{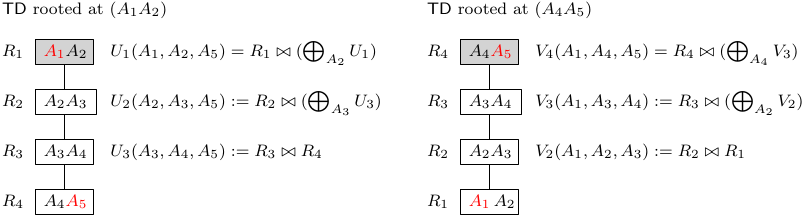}
         \caption{An illustration of two width-1 \tds~ for a line query with $k=4$. For the \td~rooted at $(A_1A_2)$, $U_3, U_2, U_1$ are the intermediate results materialized, implying a free-connex \td (left) in Figure~\ref{fig:line-4}. For the \td~rooted at $(A_4A_5)$, $V_2, V_3, V_4$ are the new relations materialized, implying a free-connex \td (right) in Figure~\ref{fig:line-4}.}
         \label{fig:td-line-4}
         \vspace{-1em}
    \end{figure}
        
    \paragraph{Semi-Joins} A tuple $t \in R_u$ for some node $u \in \nodes(\T)$ is {\em dangling} if it does not participate in any full join result, i.e., $t \notin \pi_{\chi(u)} \Join_{u' \in \nodes(\T)} R_{u'}$. The first phase removes all {\em dangling tuples} via a bottom-up and top-down pass of semi-joins along $\T$, which takes $O(N)$ time. 

\begin{wrapfigure}{r}{0.23\linewidth}
    \centering
         \vspace{-1em}
        \includegraphics[scale=1.0]{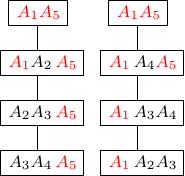}
         \vspace{-1em}
        \caption{An illustration of two free-connex \tds~ for a line query with $k=4$.}
        \label{fig:line-4}
         \vspace{-1em}
    \end{wrapfigure}

     \paragraph{Pairwise Join-Aggregation} The second phase performs joins and aggregations in a bottom-up way along $\T$. Specifically, it takes two nodes $R_u$ and $R_{u'}$ such that $u$ is a leaf and $u'$ is the parent of $u$, aggregate over non-output attributes that only appear in $\chi(u)$ but not in $\chi(u')$ by replacing $R_u$ with $\oplus_{\chi(u) - \chi(u') -\y} R_u$, and replaces $R_{u'}$ with $R_u \Join R_{u'}$. Then $R_u$ is removed, and the step repeats until only one node remains, i.e., the root node, say $r$. It will output $\oplus_{\chi(r)-\y} R_r$ as the final result. We give a running example on line queries in Figure~\ref{fig:td-line-4}. Note that for any width-1 \textsf{TD}, the intermediate join result materialized at each node implies a free-connex \td of the input query. See Figure~\ref{fig:line-4}. 
    
     The runtime of this phase is proportional to the largest intermediate join size (after dangling tuples are removed). However, this size could differ drastically on different {\em query plans}, i.e., each {\em query plan} corresponds to one width-1 \td with a particular sequence of pairwise join-aggregations. So, the runtime of the Yannakakis algorithm refers to the runtime of the {\em fastest} query plan. 

     \subsection{Line Query}   
     \label{sec:line}
\begin{figure}[t]
        \centering
        \includegraphics[scale=0.96]{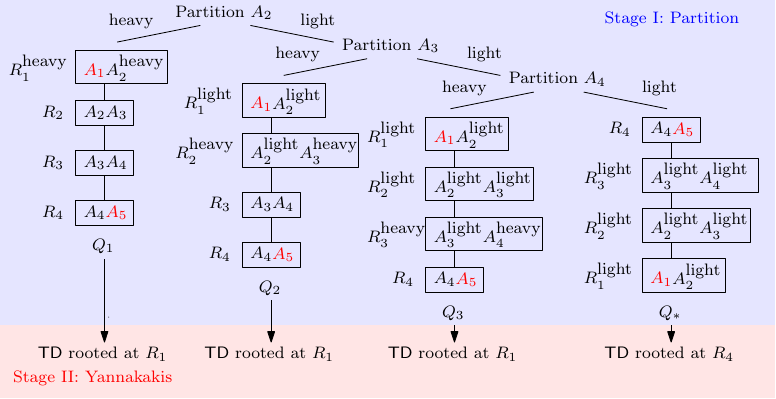}
        \vspace{-1em}
        \caption{An illustration of Algorithm~\ref{alg:line} for line query with $k=4$.}
        \label{fig:line-4-partition}
    \end{figure}

     As pointed out by \cite{hu2024fast}, there exists some instance for line queries such that every query plan materializes $\Omega(N \cdot \OUT)$ intermediate join results. 
     We exploit {\em the power of multiple query plans} to overcome this limitation. For line queries, we only consider two query plans of the Yannakakis algorithm(that correspond to two width-1 \tds): (a) one is rooted at $(A_1A_2)$; (b) one is rooted at $(A_kA_{k+1})$. See Figure~\ref{fig:td-line-4} and \ref{fig:line-4}. The high-level idea is to partition the input instance into a set of sub-instances and then choose one of two plans for each sub-instance. To determine which plan to use, we need a more fine-grained analysis of data statistics. For example, if the effective domain of $A_{k+1}$ is small, we choose the one rooted at $(A_1A_2)$; and if that of $A_1$ is small, we choose the one rooted at $(A_kA_{k+1})$. Our algorithm consists of two stages. In {\bf Stage I}, we compute the data statistics and partition the input instance. In {\bf Stage II}, we choose different query plans for each sub-instance, compute their result by the Yannakakis algorithm, and aggregate all sub-queries. The pseudocode is given in Algorithm~\ref{alg:line}. See a running example of Algorithm~\ref{alg:line} on line queries with $k=4$ in Figure~\ref{fig:line-4-partition}. 
     
     We now assume a parameter $\tOUT$ is known such that $\tOUT \le \OUT \le c \cdot \tOUT$ for some constant $c$. In  Appendix~\ref{appendix:output}, this assumption can be removed without increasing the complexity asymptotically. 
    
    \begin{algorithm}[t]
    \caption{{\sc Line}$\left(\line, \R\right)$}  
     \label{alg:line} 
     \ForEach{$i \in [k-1]$}{
        \lIf{$i=1$}{$T_1(A_1,A_2) \gets R_1(A_1,A_2)$}
        \lElse{$\displaystyle{T_i(A_1,A_{i+1}) \gets \ \bigoplus_{A_i} S_{i-1}(A_1,A_i) \Join R_i\left(A_i, A_{i+1}\right)}$}
        $A^\heavy_{i+1}\ \gets\ \left\{a \in \dom(A_{i+1}): \left|\sigma_{A_{i+1}=a} T_i\right| > \sqrt{\tOUT}\right\}$\;
        $A^\light_{i+1}\ \gets\ \left\{a \in \dom(A_{i+1}): 1 \le \left|\sigma_{A_{i+1}=a} T_i\right| \le \sqrt{\tOUT}\right\}$\;
        $R^\heavy_{i}, R^\light_{i} \gets R_{i} \ltimes A^\heavy_{i+1}, \ R_{i} \ltimes A^\light_{i+1}$\;
        $S_i(A_1,A_{i+1}) \ \gets\ T_i(A_1,A_{i+1})  \ltimes A^\light_{i+1}$\;
        $\displaystyle{\Q_i\ \gets\displaystyle{\bigoplus_{A_2,A_3,\cdots, A_k} \left(\Join_{j\in [i-1]} R^\light_j\right) \Join R^\heavy_{i}  \Join \left(\Join_{j=i+1}^k R_j\right)}}$ with a \td\ rooted at $R_1$\;}
        $\displaystyle{\Q_*\gets\bigoplus_{A_2,A_3,\cdots,A_k} \left(\Join_{j \in [k-1]} R^\light_j\right) \Join R_k}$ with a \td\ rooted at $R_k$\;
        \Return $\Q_1 \oplus \Q_2 \oplus \cdots \oplus \Q_{k-1} \oplus \Q_*$\;
    \end{algorithm}

    \smallskip
    \noindent{\bf Stage I: Partition.} We first compute some data statistics. For each value $a \in \dom(A_2)$, we define its {\em degree} as the number of tuples from $R_1$ displaying $a$ in $A_2$, i.e., $\triangle(a) = |\pi_{A_1}\sigma_{A_2 = a} R_1|$. A value $a\in\dom(A_2)$ is {\em heavy} if $\triangle(a)>\sqrt{\tOUT}$, and {\em light} otherwise. Let $A^\heavy_2, A^\light_2$ be the set of heavy and light values in $A_2$. Let $R^\heavy_1 = R_1 \ltimes A^\heavy_2$ and $R^\light_1= R_1 \ltimes A^\light_2$ be the set of heavy and light tuples in $R_1$ respectively. Below, we partition relations by ordering $R_2,R_3,\cdots,R_k$. 
    
    Suppose we are done with $R_1,R_2,\cdots,R_{i-1}$. We next partition values in $\dom(A_{i+1})$ that can be joined with any value in $\dom(A_1)$ via $R^\light_1, R^\light_2, \cdots, R^\light_{i-1}$.

    Let $\displaystyle{\triangle(a) = \left|\pi_{A_1} \left\{\left(\Join_{j \in [i-1]} R^\light_j \right) \Join (\sigma_{A_{i+1} = a
    } R_i)\right\} \right|}$ be the {\em degree} of each value $a \in \dom(A_{i+1})$. 
    A value $a \in \dom(A_{i+1})$ is {\em heavy} if $\triangle\left(a\right) > \sqrt{\tOUT}$, and {\em light} otherwise. Let $A^\heavy_{i+1}, A^\light_{i+1}$ be the set of heavy, light values in $A_i$. 
    Then, $R^\heavy_i = R_i \ltimes A^\heavy_{i+1}$ and $R^\light_i = R_i \ltimes A^\light_{i+1}$. Note that some values in $A_{i+1}$ may be undefined, as well as some tuples in $R_i$.
    
    We compute $\triangle(\cdot)$ as described in Algorithm~\ref{alg:line}. For simplicity, we define two intermediate relations:
    \begin{align*}
        T_i\left(A_1,A_{i+1}\right)  = \bigoplus_{A_2,A_3,\cdots, A_{i}}\left(\Join_{j \in [i-1]} R^\light_j \right) \Join R_i, \textrm{ and }
        S_i(A_1,A_{i+1})  = \bigoplus_{A_2,A_3,\cdots, A_{i}} \left(\Join_{j \in [i]} R^\light_j \right)
    \end{align*}
    and recursively compute them as follows (with $T_1 = R_1$ in line 2):
    \begin{align*}
        T_{i}\left(A_1,A_{i+1}\right) &= \bigoplus_{A_i} S_{i-1}\left(A_1,A_i\right) \Join R_i(A_i, A_{i+1});  \ \ \ \textrm{(line 3)} \\
        S_i\left(A_1,A_{i+1}\right) &= T_{i}\left(A_1,A_{i+1}\right) \ltimes A^\light_{i+1}; \ \ \ \textrm{(line 7)}
    \end{align*}

    \noindent Once we have computed $T_i$ for $i \in [k-1]$. We can identify the heavy and light values in $A_{i+1}$, i.e., $A^\heavy_{i+1}$ and $A^\light_{i+1}$ (lines 4-5). We can use $A^\heavy_{i+1}, A^\light_{i+1}$ to partition $R_i$ into $R^\heavy_i, R^\light_i$ (line 6). Then, $S_i$ can be computed based on $T_i$ and $A^\light_{i+1}$. Furthermore, $T_{i+1}$ can be computed based on $S_i$ and $R_{i+1}$. 

    We partition $\line$ into $k$ sub-instances: $\displaystyle{\Q_{i} =  \displaystyle{\bigoplus_{A_2,A_3,\cdots, A_k}  \left(\Join_{j\in [i-1]} R^\light_j\right) \Join R^\heavy_{i} \Join \left(\Join_{j=i+1}^k R_j\right)}}$
    for each $i \in [k-1]$, and $\displaystyle{\Q_{*} = \bigoplus_{A_2,A_3,\cdots, A_k}\left(\Join_{j\in [k-1]} R^\light_j\right) \Join R_k}$.

    \paragraph{Stage II: Yannakakis} For any $i \in [k-1]$, we compute $\Q_i$ by invoking the Yannakakis algorithm over the \td (a). For $\Q_*$, we compute $\bigoplus_{A_k} S_{k-1}(A_1,A_k) \Join R_{k}(A_k, A_{k+1})$. It is equivalent to invoking the Yannakakis algorithm over the \td (b). Finally, we aggregate the results of all sub-queries.

    \paragraph{Analysis}
    In {\bf Stage I}, consider any $i \in \{2,3,\cdots,k\}$. As there are $N$ tuples in $R_i$, and each of them can be joined with at most $\sqrt{\tOUT}$ tuples in $S_{i-1}$, $T_{i}$ can be computed within $O\left(N \cdot \sqrt{\OUT}\right)$ time. Also, $|T_{i}|=O\left(N \cdot \sqrt{\OUT}\right)$. The cost of computing $S_i$ is $O(|T_i|) = O\left(N \cdot \sqrt{\OUT}\right)$. In {\bf Stage II}, for $\Q_i$, each value in the effective domain of $A_{k+1}$ can be joined with at least $\sqrt{\tOUT}$ values in $A_1$, implied by $A^\heavy_{i+1}$. As there are $\OUT$ results in total, the effective domain size of $A_{k+1}$ is $O\left(\sqrt{\OUT}\right)$. Hence, the number of intermediate join results materialized for each node can be bounded by 
    $O\left(N \cdot \sqrt{\OUT}\right)$.
    For $\Q_*$, the total number of intermediate join results is $O\left(N \cdot \sqrt{\OUT}\right)$, as there are $N$ tuples in $R_{k}$ and each of them can be joined with at most $\sqrt{\tOUT}$ tuples in $S_{k-1}$. Hence, this step takes $O\left(N \cdot \sqrt{\OUT}\right)$ time in total. Finally, as each sub-query produces at most $\OUT$ results, the total number of results aggregated is $O(\OUT)$. Hence, this step takes $O(\OUT)$ time. Putting everything together, we obtain an algorithm for line queries with time complexity $O\left(N \cdot \sqrt{\OUT}\right)$.

    \begin{theorem}
        \label{the:line}
        For a line query $\line$ and an arbitrary instance $\R$ of input size $N$ and output size $\OUT$, the query result $\Q(\R)$ can be computed in $O\left(N\cdot \sqrt{\OUT}\right)$ time.
    \end{theorem}

    \section{Acyclic Queries}
    \label{sec:acyclic}
    After obtaining some high-level ideas behind our hybrid strategy for line queries, we are now ready to move to general acyclic queries. But, there are several challenging questions in front of us:
    \begin{itemize}[leftmargin=*]
        \item \textit{How to relate free-connex fractional hypertree width to the structure of an acyclic query?}
        \item \textit{What optimal condition is required to run the Yannakakis algorithm within the targeted time?}
        \item \textit{How to partition the input instance into a set of sub-instances satisfying the optimal condition?}
    \end{itemize}
    We will answer these critical questions step by step. All missing materials are given in Appendix~\ref{appendix:acyclic}. 
    
    \subsection{Outline of Our Algorithm}
    \label{sec:framework}
     The procedural definition of free-connex fractional hypertree width in Corollary~\ref{cor:recursive} essentially implies the outline of our algorithm. As described in Algorithm~\ref{alg:meta}, given an acyclic query $\Q$ and instance $\R$, after removing dangling tuples (line 1), we compute the results of each $\exists$-connected subquery separately (line 4-6) and finally combine their results via joins (line 7). From now on, we

    \begin{algorithm}[t]\caption{\textsc{AcyclicJoinAggregate}$(\Q, \R)$}
    \label{alg:meta}
    Remove dangling tuples in $\R$\;
    $\Q_1,\Q_2,\cdots, \Q_h \gets$ $\exists$-connected subqueries of $\Q$\; 
    \lForEach{$i \in [h]$}{$\R_i \gets$ sub-instance for $\Q_i$} 
    \ForEach{$i \in [h]$}{
        $(\Q'_i,\R'_i)  \gets \textsc{Cleanse} \textrm{ and }\textsc{Separate}(\Q_i,\R_i)$\;
        $\mathcal{S}_i \gets \textsc{HybridYannakakis}(\Q'_i, \R'_i)$\;}
        \Return $\Join_{i \in [h]} \mathcal{S}_i$ by Yannakakis algorithm~\cite{yannakakis1981algorithms}\;
    \end{algorithm}
    \begin{algorithm}[t]
    \caption{$\textsc{Separate}(\Q=(\V,\E,\y),\R)$}
    \label{alg:separate}
    \SetKwInOut{Input}{Input}
    \SetKwInOut{Output}{Output}
    
    $\E' \gets$ a set of $\fnfhtw$ relations with $\displaystyle{\y \subseteq \bigcup_{e \in \E'} e}$\;
    $\kappa \gets$ an assignment from $\y$ to $\E'$\;
    \ForEach{$A \in \y$ with $|\E_A| > 1$}{
        $x_A \gets$ an attribute not appearing $\V$\;
        $\V \gets \V \cup \{x_A\}$, $\kappa(A) \gets \kappa(A) \cup \{x_A\}$, $\y \gets \y \cup \{x_A\} - \{A\}$\;
        \lForEach{tuple $t \in R_{\kappa(A)}$}{$\pi_{x_A} t \gets \pi_A t$}
    }
    \ForEach{$e \in \E$ s.t. $\not\exists e' \in \E$ with $e -\y \subseteq e'$}{
        $x_e \gets $ an attribute not appearing in $\V$\;
        $e'' \gets \{x_e\} \cup (e\cap \y)$\;
        $\V \gets \V \cup \{x_e\}$, $\y \gets \y \cup \{x_e\} - (e\cap \y)$, $\E \gets \E \cup \{e''\}$\;
        $R_{e''} \gets \emptyset$\;
        \ForEach{$t \in \pi_{e \cap e''} R_e$}{
            $t' \gets$ a tuple over attributes $e''$ with $\pi_{e \cap e''} t' = \pi_{x_e} t' = t$ and $w(t') = \mathbf{1}$\; 
            $R_{e''} \gets R_{e''} \cup \{t'\}$\;
        }
        $\R \gets \R \cup \{R_{e''}\}$\;
    }
    \Return Updated $\Q$ and $\R$\;
    \end{algorithm}
    
    \noindent shrink our scope to $\exists$-connected queries.  Consider an arbitrary sub-query $\Q_i$ and instance $\R_i$. If $\Q_i$ is not cleansed, we apply the {\sc Cleanse} procedure to get a cleanse version $\Q'_i$ of $\Q_i$ and an updated instance $\R'$ of input size $O(|\R|)$ such that $\Q_i(\R_i) = \Q'_i(\R'_i)$, which takes $O(N)$ time. Then, we can further shrink our scope to $\exists$-connected and cleansed queries.
    In Section~\ref{sec:separate}, we propose the {\sc Separate} procedure that transforms any query falling into this case into a {\em separated} one with the same free-connex fractional hypertree width (Theorem~\ref{the:separate}). Finally, we can focus on separated acyclic queries thanks to these helper procedures. In Section~\ref{sec:separated-structure} and Section~\ref{sec:separated-acyclic-algorithm}, we further investigate the structural properties of separated acyclic queries and show our final output-optimal algorithm (Theorem~\ref{the:up-separated}).  

     \begin{theorem}
    \label{the:separate}
       Any $\exists$-connected and cleansed acyclic query $\Q$ and an instance $\R$ of input size $N$, can be transformed into a separated acyclic query $\Q'$ and an instance $\R'$ of input size $O(N)$ within $O(N)$ time, such that $\fnfhtw(\Q) = \fnfhtw(\Q')$ and $\Q(\R) = \Q'(\R')$.
    \end{theorem}
    
    \begin{theorem}
        \label{the:up-separated}
        For any separated acyclic query $\Q$, and an instance $\R$ of input size $N$ and output size $\OUT$, $\Q(\R)$ can be computed in $\displaystyle{O\left(N \cdot \OUT^{1-\frac{1}{\fnfhtw(\Q)}}\right)}$ 
        time. 
    \end{theorem}

    We now briefly analyze the complexity of Algorithm~\ref{alg:meta}. Both {\sc Cleanse} and {\sc Separate} take $O(N)$ time. Each invocation of {\sc HybridYannakakis} takes $O\left(N \cdot \left|\Q'_i\left(\R'_i\right)\right|^{1- \frac{1}{\fnfhtw(\Q'_i)}}\right)$ time. By Lemma~\ref{lem:out-width-cleansed} and Theorem~\ref{the:separate}, $\fnfhtw\left(\Q'_i\right) = \fnfhtw(\Q_i) \le \fnfhtw(\Q)$. When dangling tuples are removed, $\left|\Q'_i\left(\R'_i\right)\right| \le |\Q(\R)|$. Together, we get the desired complexity $O\left(N \cdot \OUT^{1-\frac{1}{\fnfhtw(\Q)}}+\OUT\right)$.

    \subsection{\textsc{Cleanse} Procedure}
    \label{sec:cleanse} As shown in Algorithm~\ref{alg:cleanse}, this procedure takes as input an acyclic query $\Q$ with an instance $\R$ of input size $N$, and outputs the cleansed version $\Q'$ with an updated instance $\R'$ of input size $O(N)$, such that $\Q(\R) = \Q'(\R')$. It starts with removing all dangling tuples from $\R$, and then iteratively applies the following two steps: (line 3-7) removes a unique non-output attribute $B \in \V - \y$ (suppose $\E_B = \{e\})$ and aggregate $R_e$ over $B$;  
    or (line 8-10) removes relation $e \in \E$ if there is another relation $e' \in \E$ with $e \subseteq e'$ and update $R_{e'}$ by $R_{e'} \Join R_e$, i.e., update the annotation of each tuple $t \in R_{e'}$ by $\displaystyle{w(t) \otimes w\left(\pi_{e} t\right)}$.

    \begin{algorithm}[t]
    \caption{$\textsc{Cleanse}(\Q=(\V,\E,\y),\R)$}
    \label{alg:cleanse}
    \SetKwInOut{Input}{Input}
    \SetKwInOut{Output}{Output}
    \While{$(\V,\E,\y)$ is not cleansed}{
        \If{$\exists B \in (\V - \y)$ s.t. $|\E_B| =1$, say $\E_B = \{e\}$}{
            $R_e \gets \oplus_B R_{e}$, $e \gets e - \{B\}$, $\V \gets \V - \{B\}$\; 
        }
        \If{$\exists e,e' \in \E$ s.t. $e \subseteq e'$}{
            $R_{e'} \gets R_{e'} \Join R_{e}$, $\E \gets \E- \{e\}$\;
        }
    }
    \Return Updated $\Q = (\V,\E,\y)$ and $\R$\;
    \end{algorithm}

    \subsection{\textsc{Separate} Procedure}
    \label{sec:separate}
    As shown in Algorithm~\ref{alg:separate}, the {\sc Separate} procedure takes as input an $\exists$-connected and cleansed acyclic query $\Q$ and an instance $\R$.
    We start with joint output attributes (lines 1-6).
    As proved in Lemma~\ref{lem:assignment}, it is always feasible to find a subset $\E' \subseteq \E$ of $\fnfhtw$ relations that contain all output attributes (line 1). For each attribute $A \in \y$, we pick an arbitrary relation $e \in \E'$ with $A \in e$ and {\em assign} $A$ to $e$. Let $\kappa$ be such an {\em assignment} (line 2). For each joint output attribute $A \in \y$, we add a unique output attribute $x_A$ to relation $\kappa(A)$ and ``turn'' $A$ into non-output (line 5). To preserve the equivalence of query results, we force a one-to-one mapping between $\dom(A)$ and {$\dom(x_A)$} (line 6).  The annotations of all tuples stay the same. Applying this step to every joint output attribute, we are left with a query where the set of unique attributes is exactly the set of output attributes. 
    
    We next consider any relation $e \in \E$ with $e \cap \y \neq \emptyset$, such that there is no other relation $e' \in \E$ with $e - \y \subseteq e'$ (lines 7-16). We add another relation $e''$ containing a unique output attribute $x_e$ and all output attributes in $e$, and then ``turn'' all output attributes in $e$ into non-output (line 10). To preserve the equivalence of the query results, we force a one-to-one mapping between $\dom(e \cap e'')$ and $\dom(x_e)$, and set the annotation of each tuple in $R_{e''}$ as $\mathbf{1}$ (line 12-15). It is easy to see that the number of relations containing (unique) output attributes is $\fnfhtw$.
    
    \begin{wrapfigure}{r}{0.25\linewidth}
        \centering
        \includegraphics[scale=1.3]{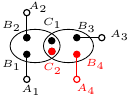}
        \vspace{-0.8em}
        \caption{An illustration of the {\sc Separate} procedure on $\Q$ in Figure~\ref{fig:Q3}.}
        \label{fig:Q3-separate}
        \vspace{-2.2em}
    \end{wrapfigure}

    \noindent \quad
    \vspace{-1.2em}
     \begin{example}
        \label{exp:separate} \ We continue with Example~\ref{exp:width} where $\Q$ is not separated. Suppose $\E' = \{R_1,R_2,R_3,R_6\}$. As $C_2$ is not unique, we add a unique output attribute $B_4$ to $R_6$ and turn $C_2$ into non-output. As no relation contains all non-output attributes of $R_6$, we add a relation $R_4(A_4, B_4)$ with a unique output attribute $A_4$ and turn $B_4$ into non-output. The resulting query $\Q'  =  \displaystyle{\bigoplus_{B_1,B_2,B_3,B_4,C_1,C_2}}  R_1(A_1,B_1) \Join R_2(A_2,B_2)\Join R_3(A_3,B_3) \Join R_4(A_4,B_4) \Join R_5(B_1,B_2,C_1,C_2) \Join R_6(B_3,B_4,C_1,C_2)$ is separated.
    \end{example}
    
    \subsection{Structural Properties of Separated Acyclic Queries}
    \label{sec:separated-structure}
    We introduce a nice structural property of separated acyclic queries regarding \fnfhtw. For each relation $e$ containing output attribute(s), there is another relation $e' \in \E$ containing all joint (non-output) attributes of $e$. Implied by the GYO reduction~\cite{abiteboul1995foundations}, we can put $e$ as a child of $e'$ in a width-1 \textsf{TD}. Applying this argument to every such relation, we obtain:     
    
     \begin{lemma}
        \label{lem:leaves} \ 
        Any separated acyclic query $\Q$ has a width-1 \td $(\T,\chi)$ such that there is a one-to-one correspondence between the leaf nodes of $\T$ and relations containing output attributes.
    \end{lemma}

      \begin{wrapfigure}{r}{0.34\linewidth}
    \centering
    \vspace{-1em}
    \includegraphics[scale=1]{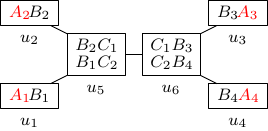}
    \vspace{-1em}
    \caption{A separated width-1 \td  for the query in Figure~\ref{fig:Q3-separate}. Nodes $u_1$, $u_2$, $u_3$ and $u_4$ are leaf nodes. All output/unique attributes are highlighted in red. $\mathcal{N}_{u_5} = \{u_1,u_2,u_6\}$. Removing edge $\{u_5,u_6\}$ leads to two sub-trees, where $\T_{u_5,u_6}$ contains nodes $u_1,u_2,u_5$ and $\T_{u_6,u_5}$ contains nodes $u_3,u_4,u_6$. $\Q_{u_5,u_6}=\bigoplus_{B_1,B_2} R_1(A_1,B_1) \Join R_2(A_2,B_2) \Join R_3(B_1,B_2,C_1,C_2)$. And, $\phi(u_5,u_6) = \phi(u_6,u_5) = \frac{1}{2}$ as nodes $u_1,u_2$ in $\T_{u_5,u_6}$ contains output attributes, and nodes $u_3,u_4$ in $\T_{u_6,u_5}$ contains output attributes. }
    \label{fig:join-tree}
    \vspace{-2em}
    \end{wrapfigure}
    
    $(\T,\chi)$ is called a {\em separated \td} for $\Q$. See Figure~\ref{fig:join-tree}. A node $u_1 \in \nodes(\T)$ is a {\em neighbor} to node $u_2 \in \nodes(\T)$ if an edge exists between $u_1,u_2$ in $\T$. Let $\mathcal{N}_{u}$ denote the set of neighbors of $u$ in $\T$. A node $u$ is a {\em leaf} if $\left|\mathcal{N}_{u}\right|=1$. The number of leaf nodes in $\T$ is exactly $\fnfhtw(\Q)$. 
    For a pair of incident nodes $u_1,u_2$, we use $\{u_1,u_2\}$ to denote the undirected edge between them, use $(u_1,u_2)$ (resp. $(u_2,u_1)$) to denote the directed edge from $u_1$ to $u_2$ (resp. from $u_2$ to $u_1$). Removing edge $\{u_1,u_2\}$ separates $\T$ into two connected subtrees $\T_{u_1,u_2}$ and $\T_{u_2,u_1}$, which contains $u_1$ and $u_2$ separately.  Let $\L_{u_1,u_2}$ be the set of leaf nodes in $\T_{u_1,u_2}$. For simplicity, we use $\phi_{u_1,u_2} = \frac{|\L_{u_1,u_2}|}{\fnfhtw(\Q)}$ to measure the {\em fraction} of the number of relations containing output attributes in $\T_{u_1,u_2}$, which will be used for partitioning instances. 
    
    Similar to the Yannakakis algorithm, given an instance $\R$ of $\Q$, we define a derived relation $R_u$ for each node $u \in \nodes(\T)$ as follows: if there exists some relation $e \in \E$ with $e = \chi(u)$, $R_u := R_e$; otherwise, $R_u := \Join_{e \in \E} \pi_{\chi(u)} R_e$ in which each tuple $t \in R_u$ has its annotation $w(t)=1$. 
    Each subtree $\T_{u_1,u_2}$ derives a sub-query 
    $\displaystyle{\Q_{u_1,u_2} :=\bigoplus_{(\V - \y) - (\chi(u_1) \cap \chi(u_2))} \underset{u \in \nodes\left(\T_{u_1,u_2}\right)}{\Join} R_u}$,  
    i.e., aggregates over all non-output attributes except the join attributes between $u_1$ and $u_2$. The intuition is that these join attributes will appear in some other relation in $\T - \T_{u_1,u_2}$, and they should be kept in the subsequent computation.

    \subsection{Our Algorithm for Separated Acyclic Queries}
    \label{sec:separated-acyclic-algorithm}
    We next exploit data statistics of input instances for which the runtime of the Yannakakis algorithm can be bounded by Theorem~\ref{the:up-separated}, and more importantly, how to partition the input instances to meet the optimal condition. Similarly, a parameter $\tOUT$ is known such that $\tOUT \le \OUT \le c \cdot \tOUT$ for some constant $c$. In Appendix~\ref{appendix:output}, this assumption can be removed without increasing the complexity asymptotically. We introduce the notion of {\em edge label} that captures both query and data properties.

    \begin{definition}[Edge Label]
    \label{def:edge-label}
    For a separated acyclic query $\Q$ and a separated width-1 \td $(\T,\chi)$, an instance $\R$ and parameter $\tOUT$, an edge $(u_1,u_2)$ in $\T$ is {\em large} if $\left|\Q_{u_1,u_2}(\R) \ltimes {t}\right| > \tOUT^{\phi_{u_1,u_2}}$ holds for every tuple $t \in \pi_{\chi(u_1) \cap \chi(u_2)} R_{u_1}$; and {\em small} if $\left|\Q_{u_1,u_2}(\R) \ltimes {t}\right| \le \tOUT^{\phi_{u_1,u_2}}$ holds for every tuple $t \in \pi_{\chi(u_1) \cap \chi(u_2)} R_{u_1}$; and 
    {\em unlabeled} otherwise. Furthermore, a small edge $(u_1,u_2)$ is 
    {\em limited} if $\left|\bigoplus_{\chi(u_1) \cap \chi(u_2)}\Q_{u_1,u_2}(\R)\right|\le \tOUT^{\phi_{u_1,u_2}}$. 
    \end{definition}

    \subsubsection{Optimal Condition of Yannakakis Algorithm}
    
    Now, we can see a natural connection between the Yannakakis algorithm and edge labels: 
    \begin{lemma}
    \label{lem:materialization}
         For a separated acyclic query $\Q$ with a separated width-1 \td $(\T,\chi)$, and an instance $\R$ of input size $N$ and output size $\OUT$, if edge $(u,u')$ is small, the intermediate join result $R_{u'} \Join \Q_{u,u'}(\R)$ can be computed in $O\left(N\cdot \OUT^{\phi_{u,u'}}\right)$ time.
    \end{lemma}

    Consider an arbitrary node $u_1$ in $\T_{u,u'} - \{u\}$. The intermediate join result materialized at $u_1$ is $R_{u_1} \Join \left(\Join_{u_2} \Q_{u_2,u_1}\right)$, where $u_2$ is over all child nodes of $u_1$. Without dangling tuples, every tuple $t\in R_{u_1}$ participates in at most $\left(\Join_{u_2} \Q_{u_2,u_1}\right)$ query result in $\Q_{u_1,p_{u_1}}$; otherwise, there must exist some tuple $t \in \dom(\chi(u) \cap \chi(u'))$ that can participate in more than $\tOUT^{\phi_{u,u'}}$ query result in $\Q_{u,u'}$, contradicting the fact that edge $(u,u')$ is small. As there are at most $N$ tuples in $R_{u_1}$, the number of intermediate join results materialized at $u_1$ is at most $O\left(N \cdot \OUT^{\phi_{u,u'}}\right)$. When we move to node $u'$, the size of $R_{u'} \Join \Q_{u,u'}(\R)$ can be bounded similarly. Hence, we obtain Lemma~\ref{lem:materialization}. 
    
    If $u'$ is a leaf node with its unique neighbor as $u$, $\phi_{u,u'} =1- \frac{1}{\fnfhtw}$. Hence: 

    \begin{corollary}[Optimal Condition for Yannakakis ]
    \label{lem:base}
        For a separated acyclic query $\Q = (\V, \E,\y)$ with a separated width-1 \td $(\T,\chi)$, and instance $\R$ of input size $N$ and output size $\OUT$, if there is a leaf node $u' \in \nodes(\T)$ (say incident to $u$) such that  edge $(u,u')$ is small,
        then $\Q(\R)$ can be computed in $O\left(N \cdot \OUT^{1-\frac{1}{\fnfhtw(\Q)}}\right)$ time.
    \end{corollary}

    \subsubsection{Partition Separated Acyclic Queries}
    \label{sec:partition}
    The input instance does not necessarily meet the optimal condition above, so we cannot apply the Yannakakis algorithm directly. Even worse, it is unknown how to efficiently {\em decide} the label of each edge since the definition of edge labeling is based on the sub-query, which may be too expensive to compute. Hence, the most technical step is to find some efficient ordering to label edges and iteratively split out a sub-instance once it meets the optimal condition in Corollary~\ref{lem:base}.

      \begin{algorithm}[t]
    \caption{\textsc{HybridYannakakis}
    $\left(\Q = (\V,\E,\y),\R\right)$}
    \label{alg:partition}
        $(\T,\chi) \gets$ a separated width-1 \td of $\Q$ with all edges of $\T$ non-labeled\;
        \ForEach{$u \in \nodes(\T)$}{\lIf{$\exists e \in \E$ with $e = \chi(u)$}{$\displaystyle{R_u:= R_e}$}
        \lElse{$R_u:=\Join_{e \in \E} \pi_{e \cap \chi(u)} R_e$ with $w(t)=1$ for each tuple $t \in R_u$}}
        $\mathcal{P} \gets \{(\T,\{R_u: u \in \nodes(\T)\})\}$,  $\mathcal{S} \gets \mathbf{0}$\;
        \While{$\mathcal{P} \neq \emptyset$}{
            $(\T',\R')\gets $ an arbitrary pair in $\mathcal{P}$\;
            \While{$\exists$ non-labeled $(u_1,u_2)$ in $\T'$ such that $(u_3,u_1)$ is limited for each $u_3 \in \mathcal{N}_{u_1} - \{u_2\}$}{
                    Label edge $(u_1,u_2)$ as limited\;
                } 
            \If{there is a leaf node $u_1$ with small (or limited) edge $(*,u_1)$ in $\T'$}{
                $\mathcal{S} \gets \mathcal{S} \oplus \Q(\R')$ for computing $\Q(\R')$ via  Yannakakis algorithm on $\T$ rooted at $u_1$\;
            }
            \Else{
                $(u_1,u_2) \gets$ a non-labeled edge in $\T'$ s.t. $(u_3,u_1)$ is small for each $u_3 \in \mathcal{N}_{u_1} - \{u_2\}$\;
                Compute $\Q_{u_1,u_2}(\R')$ by the Yannakakis algorithm on $\T_{u_1,u_2}$ rooted at $u_1$\;}
                $\mathcal{L} \gets \left\{t \in \dom(\chi(u_1) \cap \chi(u_2)): \left|\sigma_{\chi(u_1) \cap \chi(u_2) = t} \Q_{u_1,u_2}(\R')\right| > \tOUT^{\phi_{u_1,u_2}}\right\}$\;
                $\T'_1\gets \T'$ with  $(u_1,u_2)$, $(u_2,u_1)$ labeled as large, limited separately\;
                $\T'_2 \gets \T'$ with  $(u_1,u_2)$ labeled as small\;
                $\mathcal{P}  \gets \mathcal{P} \cup \left\{\left(\T'_1, \R' - \{R_{u_1}\}  \cup  \{R_{u_1} \ltimes \mathcal{L}\}\right), \left(\T'_2,\R'- \{R_{u_1}\}  \cup  \{R_{u_1} \triangleright \mathcal{L}\}\right)\right\}$\;
    
            $\mathcal{P} \gets \mathcal{P} - \{(\T',\R')\}$\;
            }
    \Return $\mathcal{S}$\;
    \end{algorithm}

    As described in Algorithm~\ref{alg:partition}, let $(\T,\chi)$ be a separated width-1 \td for $\Q$ {(line 1)}, with all edges unlabeled initially, and we compute the derived instance $\R_\T$ for $(\T,\chi)$ (line 2-3). We put $(\T,\R_\T)$ into a candidate set $\mathcal{P}$ of instances to be partitioned {(line 2)}. In general, consider an arbitrary pair $(\T',\R') \in \mathcal{P}$. From Lemma~\ref{lem:limited-imply-limited}, we apply {\em limited-imply-limited} rule to infer edge labels (lines 8-9). If it meets the optimal condition, we compute the result immediately {(line 10-11)} and remove it from $\mathcal{P}$ (line 19). Otherwise, we further partition it (lines 12-18). More specially, we pick a non-labeled edge $(u_1,u_2)$ such that every other incoming edge to $u_1$ is small, i.e., edge $(u_3, t_1)$ is small for each node $u_3 \in \mathcal{N}_{u_1} - \{u_2\}$ {(line 13)}. We compute $\Q_{u_1,u_2}(\R')$ using the Yannakakis algorithm along $\T_{u_1,u_2}$ rooted at $u_1$ {(line 14)}.
    A tuple $t \in \dom(\chi(u_1) \cap \chi(u_2))$ is {\em heavy} if $\left|\Q_{u_1,u_2}(\R') \ltimes t\right| > \tOUT^{\phi_{u_1,u_2}}$, and {\em light} otherwise. Now, we construct two sub-instances for $\R'$, which contain heavy and light tuples in $R_{u_1}$ separately {(line 18)}, and two copies of $\T'$ in which edge $(u_1, u_2)$ is further labeled as large and small separately {(line 16-17)}. By Lemma~\ref{lem:large-reverse-limited}, we can apply {\em large-reverse-limited} rule to infer $(u_2,u_1)$ as limited when $(u_1,u_2)$ is labeled as large {(line 16)}. We add these two sub-instances into $\mathcal{P}$ {(line 18)} and also remove $(\T',\R')$ from $\mathcal{P}$ {(line 19)}. We continue applying this procedure to every remaining instance in $\mathcal{P}$ until $\mathcal{P}$ becomes empty {(line 6)}.

    \begin{lemma}[Limited-Imply-Limited]
    \label{lem:limited-imply-limited}
        For any node $u_1$, if there exists a node $u_2 \in \mathcal{N}_{u_1}$ such that edge $(u_3,u_1)$ is limited for every node $u_3 \in \mathcal{N}_{u_1} - \{u_2\}$, then edge $(u_1,u_2)$ must be limited.  
    \end{lemma}
    \begin{lemma}[Large-Reverse-Limited]
    \label{lem:large-reverse-limited}
        If edge $(u_1,u_2)$ is large, then edge $(u_2,u_1)$ must be limited.
    \end{lemma}
    \vspace{-0.5em}
    \begin{example}
    \label{exp:partition} We continue the example in Figure~\ref{fig:join-tree}. Initially, all edges are unlabeled in (9.1). We start with applying line 13 to label edge $(u_1,u_5)$ since $\mathcal{N}_{u_1} - \{u_5\} = \emptyset$. In (9.2), the instance with large edge $(u_1,u_5)$ and limited edge $(u_5,u_1)$ already meets the optimal condition. The remaining instance has a small edge $(u_5,u_1)$. We can apply a similar argument to edges $(u_2,u_5)$, $(u_3,u_6)$ and $(u_4,u_6)$. In (9.3), we are left with the remaining instance with small edges $(u_1,u_5)$, $(u_2,u_5)$, $(u_3,u_6)$ and $(u_4,u_6)$. Then, we can apply line 13 to label edge $(u_5,u_6)$.

    In (9.4), for the instance with large edge $(u_5,u_6)$ and limited edge $(u_6,u_5)$, we can apply line 13 to partition both edges $(u_5,u_1)$ and $(u_5,u_2)$. Suppose we partition edge $(u_5,u_1)$ wlog. In (9.6), the instance with small edge $(u_5,u_1)$ already meets the optimal condition. In (9.7), we are left with an instance with large edge $(u_5,u_1)$ and limited edge $(u_1,u_5)$. We can apply the limited-imply-limited rule to infer edge $(u_5,u_2)$ as limited. This instance also meets the optimal condition.
    \begin{figure}[t]
        \centering
        \includegraphics[width=0.95\linewidth]{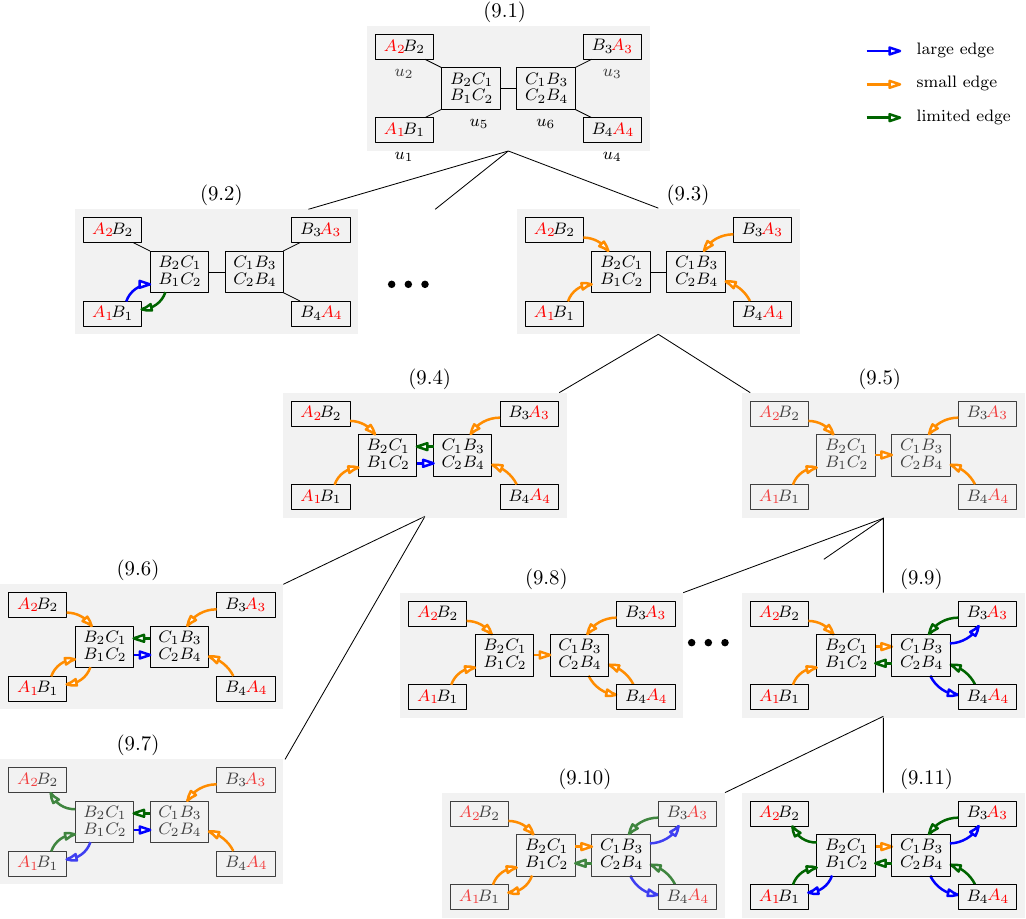}
        \vspace{-1em}
        \caption{An illustration of the partition procedure on the query in Figure~\ref{fig:join-tree}.}
        \label{fig:partition}
    \end{figure}
    
    In (9.5), for the other instance with small edge $(u_5,u_6)$, we can apply line 13 to label both edges $(u_6,u_4)$ and $(u_6,u_5)$. Suppose we label edge $(u_6,u_4)$ wlog. In (9.8), the instance with small edge $(u_6,u_4)$ meets the optimal condition. We can apply a similar argument to edge $(u_6,u_3)$. In (9.9), the remaining instance has large edges $(u_6,u_3)$ and $(u_6,u_4)$, as well as limited edges $(u_3,u_6)$ and $(u_4,u_6)$. We can apply the limited-imply-limited rule to infer edge $(u_6,u_5)$ as limited. Now, we can apply line 13 to label edges $(u_5,u_1)$ and $(u_5,u_2)$. Suppose we label edge $(u_5,u_1)$. In (9.10), the instance with small edge $(u_5,u_1)$ already meets the optimal condition. In (9.11), we are left with the instance with large edge $(u_3,u_1)$ and limited edge $(u_3,u_2)$. Then, we can apply the limited-imply-limited rule to infer edge $(u_3,u_2)$ as limited. This instance also meets the optimal condition. Now, $\mathcal{P} = \emptyset$, and we are done with the partition procedure.
    \end{example}

    The complexity of the partition procedure is stated in Lemma~\ref{lem:partition}. All other missing materials are given in Appendix~\ref{appendix:acyclic}. Combine Corollary~\ref{lem:base} and Lemma~\ref{lem:partition}, we finally obtain:

   \begin{lemma}
    \label{lem:partition}
       For a separated acyclic query $\Q$ with a separated width-1 \td $(\T,\chi)$, an arbitrary instance $\R$ can be partitioned into $O(1)$ sub-instances $\R_1,\R_2,\cdots, \R_h$ within $O\left(N \cdot \OUT^{1-\frac{1}{\fnfhtw(\Q)}}\right)$ time such that $\Q(\R) = \bigoplus_{i \in [h]} \Q(\R_i)$, and each $\R_i$ satisfies Corollary~\ref{lem:base} together with $\T$.
    \end{lemma}

    \section{Related Work}    
    
    \paragraph{Other efficient evaluation of Join-Aggregate queries} Olteanu and Z{\'a}vodn{\`y} \cite{olteanu2015size} investigated the problem of representing and computing results of conjunctive queries in a factorized form. It introduced the width $s^\uparrow(\Q)$ that captures the time to compute and size of factorized representations for conjunctive queries with arbitrary free variables. Moreover, the mapping between this notion of width and free-connex fractional hypertree width has been shown. Some follow-up works further investigated computing aggregates over factorized databases~\cite{DBLP:conf/sigmod/SchleichOC16}. More specially, evaluating the aggregate functions such as SUM, PROD, MIN, MAX, on top of the factorized representations can be done within $O(N^{s^\uparrow(Q)})$ time. Later, Abo Khmais et al. \cite{abo2016faq} went further and generalized it to arbitrary semirings and also to several semirings. More specifically, a FAQ query can be computed in $O(N^{\textrm{faqw}(\Q)} + \OUT)$ time, where $\textrm{faqw}(\Q)$ is the FAQ-width of the query.
    
    \paragraph{Yannakakis Algorithm and its extensions} Recent studies extend the Yannakakis algorithm to support different operators and scenarios, including projections~\cite{bagan2007acyclic}, unions~\cite{carmeli2019enumeration}, differences~\cite{hu2023computing}, comparisons~\cite{wang2022conjunctive}, dynamic query processing~\cite{idris17:_dynam, wang2023change}, massively parallel query processing~\cite{hu2019output, afrati2014gym} and secure query processing~\cite{wang2021secure,wang2022query}. In addition, several works ~\cite{birler2024robust, bekkers2024instance, zhao2025debunking, yang2023predicate} focus on implementing the Yannakakis algorithm efficiently within practical database systems. 

    \paragraph{Fast matrix multiplication for join-project queries}  Fast matrix multiplication has recently attracted much attention in the database community for processing join-project queries (a special class of join-aggregate queries defined on Boolean semi-ring). Suppose that computing two rectangular matrices of size $n^a \times n^b$ and $n^b \times n^c$ can be done in $O(n^{\omega(a,b,c) + o(1)})$ time. For simplicity, we use $\omega$ to denote $\omega(1,1,1)$. There are some important constants related to rectangular matrix multiplication, such as, $\alpha \le 1$ defined as the largest constant such that $\omega(1,\alpha,1)=2$, and $\mu$ is the (unique) solution to the equation $\omega(\mu, 1, 1) = 2\mu + 1$. Note that $\alpha = 1$ if and only if $\omega=2$, and the current best bounds on $\alpha$ are ${0.321334 <} \alpha \le 1$~{\cite{williams2024new}}. Moreover, $\mu = \frac{1}{2}$ if $\omega=2$, and the current best bounds on $\mu$ are $\frac{1}{2} \le \mu  {< 0.527661}$~{\cite{williams2024new}}. 
    
    Amossen and Pagh~\cite{amossen2009faster} first proposed an algorithm for the Boolean matrix multiplication by using fast matrix multiplication, which runs in $\tilde O\left(N^{\frac{2}{3}}\cdot \OUT^{\frac{2}{3}} + N^{\frac{(2-\alpha)\omega -2}{(1+\omega)(1-\alpha)}} \cdot \OUT^{\frac{2-\alpha \omega}{(1+\omega)(1-\alpha)}} +\OUT\right)$ time when $\OUT \ge N$, and $\tilde O\left(N \cdot \OUT^{\frac{\omega-1}{\omega+1}}\right)$ time when $\OUT < N$. Very recently, Abboud et al.~\cite{abboud2023time} have completely improved this to $\tilde O\left(N \cdot \OUT^{\frac{\mu}{1+\mu}} + \OUT + N^{\frac{(2+\alpha)\mu}{1+\mu}} \cdot \OUT^{\frac{1-\alpha \mu}{1+\mu}}\right)$. On the other hand, improving this result for any value of $\OUT$ is rather difficult, assuming the hardness of the {\em all-edge-triangle} problem~\cite{abboud2023time}. Many algorithms are proposed for Boolean matrix multiplication, whose complexity is also measured by the domain size of attributes; we refer readers to~\cite{abboud2023time} for details.  Deep et al.~\cite{deep2020fast} and Huang et al. \cite{huang2022density} also investigated the efficient implementation of these algorithms on sparse matrix multiplication in practice. Very recently, Hu \cite{hu2024fast} first applied fast matrix multiplication to speed up acyclic join-project queries and showed a polynomial large improvement over the combinatorial Yannakakis algorithm. As an independent line of research, fast matrix multiplication has been extensively used in the algorithm community to speed up detecting, counting, and listing subgraph patterns, such as cliques~\cite{DBLP:journals/algorithmica/AlonYZ97, patrascu2010towards, dalirrooyfard2024towards} and cycles~\cite{jin2024listing}. In addition, this technique has been used to approximately count cycles~\cite{tvetek2022approximate, censor2024fast}, k-centering in graphs~\cite{jin2025beyond}, etc.

    \section{Conclusion}
    In this paper, we established matching lower and upper bounds for computing general acyclic join-aggregate queries in an output-optimal manner, characterized by the free-connex fractional hypertree width of the queries. This result generalizes and improves upon all previously known upper and lower bounds. As a by-product, it also implies new output-sensitive algorithms for certain cyclic queries, but their optimality is still unknown. Hence, achieving output-optimal algorithms for cyclic join-aggregate queries remains an interesting open question. In addition, it is interesting to investigate join-aggregate queries defined over multiple semi-rings \cite{abo2016faq} and to explore if one can apply our new output-optimal algorithm to improve processing multiple semi-rings.

    \newpage
    \bibliographystyle{abbrv}
    \bibliography{paper}
    \appendix
    
    \newpage
    \section{Lower Bound}
    \label{appendix:lower-bound}
    In this section, we will show our lower bound. All missing materials can be found in Appendix~\ref{appendix:lower-bound}. 
    \begin{theorem}
    \label{the:lb-main}
        For an arbitrary self-join-free acyclic query $\Q= (\V,\E,\y)$, given any parameters $1 \le N$ and $\displaystyle{\OUT \le \max_{\R' \in \Re(N)}\left|\Q(\R')\right|}$, there exists an instance $\R$ of input size $N$ and output size $\OUT$ such that any semi-ring algorithm computing $\Q(\R)$ requires at least $\Omega\left(N \cdot \OUT^{1-\frac{1}{\fnfhtw(\Q)}} + \OUT \right)$ time, where $\fnfhtw(\Q)$ is the free-connex fractional hypertree widthof $\Q$, and $\Re(N)$ is the set of all instances for $\Q$ of input size $N$.
    \end{theorem}

    In~\cite{amossen2009faster}, it has been shown that for $\matrix$, and parameters $1\le N$ and $\OUT\le N^2$, there exists an instance $\R$ of input size $N$ and output size $\OUT$ such that any semiring algorithm for computing $\matrix(\R)$ requires $\Omega\left(N \cdot \sqrt{\OUT} \right)$ time. A similar argument has been extended to star queries: 
    \[\star = \bigoplus_{B} R_1(A_1, B) \Join R_2(A_2,B) \Join \cdots \Join R_k(A_k,B)\]
    which will used to establish our new lower bound for general acyclic queries.

    \begin{lemma}
    \label{lem:star}
        For $\star$ of $k$ relations, any parameters $1 \le N$ and $\OUT \le N^k$, there exists an instance $\R_\textsf{\upshape star}$ of input size $N$ and output size $\OUT$ such that any semiring algorithm for computing $\star(\R_\textsf{\upshape star})$ requires at least $\Omega\left(N \cdot \OUT^{1-\frac{1}{k}}\right)$ time. 
    \end{lemma}

    Note that Lemma~\ref{lem:star} has been proved for $\matrix$ when $k=2$ by Pagh and Stockel \cite{pagh14:_input}. Below, we simply generalized their proof to $\star$.
    
    \begin{proof}[Proof of Lemma~\ref{lem:star}]
        We construct a hard instance $\R_\textsf{star}$ of input size $N$ and output size $\OUT$ for $\star$ as follows. There are $\OUT^{\frac{1}{k}}$ distinct values in each attribute $A_i$ for $i \in [k]$. There are $\frac{N}{k \cdot \OUT^{1/k}}$ distinct values in attribute $B$. Each relation $R_i$ is a Cartesian product between $A_i$ and $B$. It can be checked that each relation contains exactly $\frac{N}{k}$ tuples, so the input size of $\R$ is exactly $N$. The query result of $\star(\R_\textsf{star})$ is all combinations of values in attributes $A_1, A_2, \cdots, A_k$, and the output size is exactly $\OUT$. On this hard instance $\R_\textsf{star}$, for each query result $(a_1,a_2,\cdots,a_k) \in \dom(A_1) \times \dom(A_2) \times \cdots \times \dom(A_k)$, its annotation is defined as $\displaystyle{\bigoplus_{b \in \dom(B)} \bigotimes_{i \in [k]} w(a_i, b)}$,
        which takes at least $|\dom(B)|$ operations to compute $w(a_1, a_2, \cdots, w_k,b) = \bigotimes_{i \in [k]} w(a_i, b)$ for each $b\in \dom(B)$. Note that every pair of query results do not share any common operations. 

        Similar to \cite{pagh14:_input}
        we require the algorithm to work over fields of infinite size such as real numbers. We consider each output value as a polynomial over nonzero entries of the input matrices. Again, by the Schwartz-Zippel theorem~\cite{motwani1996randomized}, two polynomials agree on all inputs if and only if they are identical. Since we are working in the semiring model, the only way to get the annotation for the query result $(a_1,a_2,\cdots,a_k)$ in an output polynomial is to directly multiply these input entries. That means that
        to compute a query result $(a_1,a_2,\cdots,a_k)$, one needs to compute a polynomial that is identical to the sum of elementary products $\displaystyle{\bigoplus_{b \in \dom(B)} \bigotimes_{i \in [k]} w(a_i, b)}$. 
        Summing over all query results, the number of operations required is at least $N \cdot \OUT^{1-\frac{1}{k}}$. Hence, any semiring algorithm requires at least  $\Omega\left(N \cdot \OUT^{1-\frac{1}{k}}\right)$ time for computing $\star(\R_\textsf{star})$. 
    \end{proof}  

    \begin{lemma}
    \label{lem:independent-S}
        For any $\exists$-connected acyclic query $\Q = (\V,\E,\y)$, there exists a subset $S \subseteq \y$ of $\fnfhtw(\Q)$ attributes such that no pair of them appear in the same relation from $\E$.
    \end{lemma}

    \begin{proof}[Proof of Lemma~\ref{lem:independent-S}]
        Note that $q[\y]$ is also acyclic. Initially, we set $S = \emptyset$. As shown in~\cite{hu2021cover}, the following greedy strategy leads to an optimal fractional edge covering for $q[\y]$ that is also integral.  It iteratively performs the following two procedures: (\romannumeral 1) removes a relation $e$ if there exists another relation $e'$ such that $e \subseteq e'$; (\romannumeral 2) if there exists a relation $e$ containing some unique attribute, we remove relation $e$ as well as all attributes in $e$, and add an arbitrary attribute in $e$ to $S$. It can be easily checked that $|S| = \fnfhtw$, and no pair of them appears in the same relation from $\E$. 
    \end{proof}

    \begin{proof}[Proof of Theorem~\ref{the:lb}] First, $\Omega(N + \OUT)$ is a trivial lower bound. Hence, it suffices to show the lower bound $\Omega\left(N \cdot \OUT^{1-\frac{1}{\fnfhtw(\Q)}}\right)$ when $\OUT < N^{\fnfhtw(\Q)}$. We will prove it via three steps:
    
    \paragraph{Step 1} Consider any $\exists$-connected and cleansed acyclic join-aggregate query $\Q=(\V,\E,\y)$. Suppose we are given parameters $1\le N$ and $\OUT < N^{\fnfhtw(\Q)}$. We show a reduction from $\Q$ to $\star$ with $\fnfhtw(\Q)$ relations. Implied by Lemma~\ref{lem:independent-S}, let $S \subseteq \y$ be the set of output attributes such that no pair appears in the same relation, and $|S| = \fnfhtw(\Q)$. Suppose we are given a hard instance $\R_\textsf{star}$ for $\star$ with $\fnfhtw(\Q)$ relations, which has input size $N$ and output size $\OUT$. We next construct an instance $\R$ for $\Q$ as follows. 
    
    Each output attribute $A \in \y - S$ contains one distinct value $\{*\}$. Each output attribute $A \in S$ contains $\OUT^{\frac{1}{\fnfhtw(\Q)}}$ distinct values. Each non-output attribute $B \in (\V - \y)$ contains $\frac{N}{|\E| \cdot \OUT^{\frac{1}{\fnfhtw(\Q)}}}$ distinct values. 
    For each relation $R_e$, $|e \cap S| \le 1$. 
    The projection of $R_e$ onto all non-output attributes contains tuples in a form of $(b_i,b_i,\cdots, b_i)$, for $i \in \left[N \cdot \OUT^{1-\frac{1}{\fnfhtw(\Q)}}\right]$. The projection of $R_e$ onto all output attributes is the full Cartesian product. Moreover, $R_{e} = \left(\pi_{\y} R_e\right) \times \left(\pi_{\V - \y} R_e\right)$. For each output attribute $A \in S$, we choose an arbitrary relation $e \in \E$ such that $A \in e$. Note that all chosen relations are also distinct. Let $\E_S$ be the set of chosen relations. We specify an arbitrary one-to-one mapping from relations in $\star$ and relations in $\E_S$, say $R_i$ corresponds to $S_i$. From our construction above, there is also a one-to-one mapping between tuples in $R_i$ and $S_i$.
    We just set the annotation of a tuple $t \in S_i$ as the same as $t' \in R_i$, if $t$ corresponds to $t'$. For every remaining relation $R_e$ in $\R$, we simply set the annotation of each tuple as $1$.  As $|e \cap S| \le 1$ for each relation $e \in \E$, it can be checked that each relation contains at most $N$ tuples. The input size of the constructed instance is $N$, and the output size is exactly $\OUT$. 
    
    It is not hard to see that $\displaystyle{\bigoplus_{\V - S} \Q(\R) = \star(\R_{\textsf{star}})}$. Any semiring algorithm that compute $\Q(\R)$ within $O\left(N \cdot \OUT^{1-\frac{1}{\fnfhtw(\Q)}}\right)$ time, can also compute $\star(\R_{\textsf{star}})$ within $O\left(N \cdot \OUT^{1-\frac{1}{\fnfhtw(\Q)}}\right)$ time. Hence, this automatically follows Lemma~\ref{lem:star}. 
    
    \paragraph{Step 2} Consider any $\exists$-connected (but not cleansed) acyclic join-aggregate query $\Q = (\V,\E,\y)$. Let $\Q' = (\V', \E',\y)$ be the cleansed version of $\Q$, and $\R'$ be the hard instance for $\Q'$ as constructed above. Each non-output attribute $B \in \V -\V'$ contains one distinct value $\{*\}$. For each $e' \in \E - \E'$, there must exist a relation $e \in \E'$ such that $e' \subseteq e$. The relation $R_{e'}$ is just a projection of $R_e$ onto attribute $e'$, where each tuple has its annotation as $1$. It is not hard to see $\displaystyle{\bigoplus_{\V- \y'} \Q'(\R') = \bigoplus_{\V -\y} \Q(\R)}$. Any semiring algorithm that can compute $\Q(\R)$ within $O\left(N \cdot \OUT^{1-\frac{1}{\fnfhtw(\Q)}}\right)$ time, can also compute $\Q'(\R')$ within $O\left(N \cdot \OUT^{1-\frac{1}{\fnfhtw(\Q)}}\right)$ time. Hence, this automatically follows {\bf Step 1}.

    \paragraph{Step 3} Consider any $\exists$-disconnected acyclic join-aggregate query $\Q = (\V,\E,\y)$. Let $\Q_1, \Q_2,\cdots, \Q_h$ be the $\exists$-connected sub-queries of $\Q$. Wlog, assume $\fnfhtw(\Q_1) = \fnfhtw(\Q)$. Suppose we are given any parameters $1\le N$ and $\OUT< N^{\fnfhtw(\Q)}$. There exists an instance $\R_1$ of input size $N$ and output size $\OUT$ such that any semiring algorithm computing $\Q_1(\R_1)$ requires $\Omega\left(N \cdot \OUT^{1-\frac{1}{\fnfhtw(\Q)}} + \OUT\right)$ time.  We can extend the sub-instance $\R_1$ to an instance $\R$ to $\Q$ as follows. Each attribute in $\V- \V_1$ contains one distinct value $\{*\}$. Consider an arbitrary relation $e \in \E-\E_1$. If $e \cap \V_1 = \emptyset$, relation $R_e$ only contains one tuple in a form of $(*,*,\cdots,*)$. Otherwise, $e \cap \V_1 \subseteq \y$. In this case, $R_e = (\pi_\y R_e) \times (\pi_{\V - \y} R_e)$. Note that each tuple in $R_e$ has its annotation as $1$. It is not hard to see $\displaystyle{\bigoplus_{\V_1 -\y_1} \Q_1(\R_1) = \bigoplus_{\V - \y} \Q(\R)}$. Any semiring algorithm that can compute $\Q(\R)$ within $O\left(N \cdot \OUT^{1-\frac{1}{\fnfhtw(\Q)}} + \OUT\right)$ time, can also compute $\Q_1(\R_1)$ within $O\left(N \cdot \OUT^{1-\frac{1}{\fnfhtw(\Q)}} + \OUT\right)$ time. Hence, this automatically follows {\bf Step 2}.
    \end{proof}

    \section{Missing Proofs in Section~\ref{sec:fnfhtw}}
    \label{appendix:fnfhtw}

    \subsection{Helper Lemmas}

    \begin{lemma}
        \label{lem:rho}
        For any query $\Q = (\V,\E,\y)$, any optimal fractional edge covering $\rho^*(\cdot)$ of $\Q$ satisfies:
        \begin{itemize}[leftmargin=*]
            \item If there exists a unique attribute $A \in \V$, say $\E_A = \{e\}$, then $\rho^*(q) = 1 + \rho^*(q[\V-e])$.
            \item  For any subsets $S_1\subseteq S_2 \subseteq \V$ of attributes, $\rho^*(q[S_1]) \le \rho^*(q[S_2])$.
        \end{itemize}
    \end{lemma}
    \begin{lemma}
    \label{lem:root}
        For any $\exists$-connected query $\Q = (\V,\E,\y)$ and any \td $(\T,\chi)\in \ftd(\Q)$, there exists a node $u \in \nodes(\T)$ such that $\y = \chi(u)$.
    \end{lemma}

     \begin{proof}[Proof of Lemma~\ref{lem:root}]
        As $\Q$ is $\exists$-connected, $e - \y \neq \emptyset$ holds for every relation $e \in \E$; otherwise, $e$ becomes a singular $\exists$-connected subquery of $\Q$. Let $(\T,\chi)$ be an arbitrary free-connex \td of $\Q$, with $S \subseteq \nodes(\T)$ as the connex of $\T$. Let $\T_1,\T_2,\cdots, \T_\ell$ be the set of connected subtrees for some $\ell \in \mathbb{Z}^+$, after removing $S$ from $\T$. For simplicity, we define $\E_i =\{e \in \E: \exists u \in \nodes(\T_i), e \subseteq \chi(u)\}$. It is easy to see that $(\E_1,\E_2, \cdots, \E_\ell)$ forms a partition of $\E$. On other hand, $\E = \bigcup_{i \in [\ell]}\E_i$. On the other hand, $\E_i \cap \E_j = \emptyset$ for any $i,j \in [\ell]$ with $i \neq j$. Suppose not, let $e \in \E_i \cap \E_j$. As $e - \y \neq \emptyset$ and $e'\subseteq \y$ for every node $e' \in \nodes(S)$, the set of nodes containing non-output attributes in $e-\y$ does not form a connected subtree, contradicting the fact that $(\T,\chi)$ is a valid \textsf{TD}.
        For any $i \in [\ell]$, relations in $\E_i$ must be $\exists$-disconnected from remaining relations. As $\Q$ is $\exists$-connected, there must exist some $i \in [\ell]$ with $\E_i = \E$ and $\E_j = \emptyset$ for any other $j$. Let $u \in \nodes(S)$ be the unique node sharing an edge with some node in $\T_i$. As $\bigcup_{u_1 \in \nodes(S)} \chi(u) = \y$, and $\chi(u_1) \cap \chi(u_2) \subseteq \chi(u)$ for ever pairs of nodes $u_1 \in \nodes(S), u_2 \in \nodes(\T_i)$, $\y \subseteq \chi(u)$. Together with $\chi(u) \subseteq \y$, $\y = \chi(u)$.
    \end{proof}

     \begin{proof}[Proof of Lemma~\ref{lem:out-width-cleansed}]
    \underline{\em Direction $\fnfhtw(\Q) \le \fnfhtw(\Q')$.} Consider any \td $(\T',\chi') \in \ftd(\Q')$. We next construct a \td $(\T,\chi)$ for $\Q$ with the same width. We consider the post-ordering of relations $e_1,e_2,\cdots, e_\ell$ removed by the cleanse process. Suppose $e_1$ is removed due to some relation $e$ (fully containing all its attributes). Let $u \in \nodes(\T')$ be a node such that $e \subseteq \chi(u)$. We add a new node $u_1$ as a child of $u$ and set $\chi(u_1) = e_1$. We move to the next relation and apply this procedure recursively. 
    As $\rho^*(q[\chi(u_1)]) =1$ for each node $u_1 \in \nodes(\T) - \nodes(\T')$, $\width(\T',\chi') = \width(\T,\chi)$.

    \underline{\em Direction $\fnfhtw(\Q) \ge \fnfhtw(\Q')$.} Consider an arbitrary \td $(\T,\chi) \in \ftd(\Q)$. Let $\Q = (\V,\E,\y)$ and $\Q'= (\V', \{e \cap \V': e\in \E\}, \y)$. It can be easily checked that $(\T,\chi)$ is also a valid \td for $\Q'$. Hence, $\ftd(\Q) \subseteq \ftd(\Q')$, and the inequality follows. 
    \end{proof}

    \begin{lemma}
    \label{lem:td-unique-attribute}
        For any query $\Q = (\V, \E,\y)$, there exists a \td $(\T,\chi)$ with the smallest width while satisfying the following property: each unique attribute $A \in \V$ only appears in one bag of $\T$.
    \end{lemma}

    \begin{proof}[Proof of Lemma~\ref{lem:td-unique-attribute}]
        Let $(\T,\chi)$ be an arbitrary \td of $\Q$ with the smallest width. We perform the following step if a unique attribute $A \in \V$ appears in more than one bag of $\T$. Consider an arbitrary node $u \in \nodes(\T)$ with $A \in \chi(u)$. We construct another mapping $\chi': \nodes(\T) \to 2^{\V}$ as follows: $\chi'(v) = \chi(v) - \{A\}$ if $v \neq u$ and $\chi'(v) = \chi(v)$ otherwise. It can be easily checked that $(\T,\chi')$ is a valid \td if $\Q$ with $\width(\T,\chi') \le \width(\T,\chi)$. We iteratively apply this argument until each unique attribute only appears in one bag of $\T$. 
    \end{proof}

    \subsection{Proof of Lemma~\ref{lem:fn-fhtw-disconnected} and Lemma~\ref{lem:fn-fhtw-connected}}
    \begin{lemma}
    \label{lem:out-width-disconnected}
         For an acyclic query $\Q$, if $\Q$ is $\exists$-disconnected with $\exists$-connected subqueries $\Q_1,\Q_2,\dots, \Q_h$, $\fnfhtw(\Q) = \max_{i \in [h]} \fnfhtw(\Q_i)$.
    \end{lemma}

     \begin{proof}[Proof of Lemma~\ref{lem:out-width-disconnected}]
        Let $(\E_1, \E_2,\cdots, \E_h)$ be the partition of $\E$ defined by the $\exists$-connectivity of $\Q$. For each $i \in [h]$, $\Q_i = (\V_i, \E_i, \y_i)$ with $\V_i = \bigcup_{e \in \E_i} e$ and $\y_i = \y \cap \V_i$.
        We first note an equivalent expression of that right-hand-side of the equation above:
        \begin{equation}
        \label{eq:decompose}\displaystyle{\max_{i \in [h]} \min_{(\T_i,\chi_i) \in \ftd(\Q_i)} \width(\T_i,\chi_i) = \min_{\substack{(\T_1,\chi_1), \cdots,(\T_h,\chi_h) \in\\ \ftd(\Q_1) \times \cdots \times \ftd(\Q_h)}} \max_{i \in [h]} \ \ \width(\T_i,\chi_i)} 
        \end{equation}
        
        \underline{\em Direction (\ref{eq:out-width}) $\le $ (\ref{eq:decompose})}. Consider an arbitrary combination $\{(\T_1,
        \chi_1), \cdots, (\T_h,
        \chi_h)\} \in \ftd(\Q_1) \times \cdots \times \ftd(\Q_h)$. We next construct a \td $(\T,
        \chi)$ for $\Q$. From Lemma~\ref{lem:root}, there exists a node $u_i \in \nodes(\T_i)$ with $\y_i = \chi(u_i)$. We root $\T_i$ at node $u_i$. We next build a width-1 $\td$ $(\T',\chi')$ for the sub-query $\Q'= (\y, \{\y_i:i\in [h]\})$ that there is a one-to-one correspondence between bags in $\T'$ and relations in $\{\y_i:i\in [h]\}$. For each $i \in [h]$, let $u \in \nodes(T')$ be the node with $\chi(u) = \y_i$. We add $\T_i$ as a child of $u$. It can be checked that $\displaystyle{\width(\T,\chi) = \max\left\{\width(\T',\chi'), \max_{i \in [h]} \width(\T_i,\chi_i)\right\} =\max_{i \in [h]} \width(\T_i,\chi_i)}$.

        \underline{\em Direction (\ref{eq:out-width}) $\ge$ (\ref{eq:decompose})}. Consider an arbitrary \td $(\T,\chi) \in \ftd(\Q)$. For each $i \in [h]$, we construct a mapping $\chi_i: \nodes(\T) \to 2^{\V_i}$ as follows. For each node $t \in 
        \nodes(\T)$, we set $\chi_i(t) = \chi(u) \cap \V_i$. $(\T,\chi_i)$ is a valid \td for $\Q_i$. Moreover, $\width(\T,\chi_i) \le \width(\T,\chi)$. For the combination $\{(\T, \chi_1), (\T, \chi_2), \cdots, (\T, \chi_h)\}$, $\displaystyle{\width(\T,\chi) \ge \max_{i \in [h]} \width(\T,\chi_i)}$. Hence, (\ref{eq:out-width}) $\ge$ (\ref{eq:decompose}) follows.
    \end{proof}

    \begin{lemma}
    \label{lem:out-width-connected}
         For an acyclic query $\Q$, if $\Q$ is $\exists$-connected, $\fnfhtw(\Q)= \rho^*\left(q[\y]\right)$, i.e., the fractional edge covering number of the sub-query induced by output attributes.
    \end{lemma}

     \begin{proof}[Proof of Lemma~\ref{lem:out-width-connected}]
    From Lemma~\ref{lem:out-width-cleansed}, if $\Q$ is not cleansed, we have $\fnfhtw(\Q) = \fnfhtw(\Q')$, where $\Q'$ is the cleansed version of $\Q$. Then, it suffices to show that for any a $\exists$-connected and cleansed query $\Q$, we have $\displaystyle{\fnfhtw(\Q)= \rho^*(q[\y])}$. 
    
    \underline{\em Direction $\displaystyle{\fnfhtw(\Q) \ge \rho^*(q[\y])}$.} Consider an arbitrary \td $(\T,\chi) \in \ftd(\Q)$. By Lemma~\ref{lem:root}, there is a node $u \in \nodes(\T)$ with $\y =\chi(u)$. Then, $\rho^*(q[\y]) = \rho^*(q[\chi(u)]) \le \width(\T,\chi)$. Applying this argument to all possible \tds~in $\ftd(\Q)$, we have $\rho^*(q[\y]) \le \fnfhtw(\Q)$.

        \underline{\em Direction $\displaystyle{\fnfhtw(\Q) \le \rho^*(q[\y])}$.}  
        It suffices to identify a \td $(\T,\chi) \in \ftd(\Q)$ with $\width(\T,\chi) = \rho^*(q[\y])$. 
        Let $(\T,\chi)$ be a width-1 
        \td of $\Q$, where each unique (output) attribute only appears in one bag of $\T$.  As $\Q$ is cleansed, a unique output attribute $A \in \y$ must exist and only appear in one bag, say $u$. 
        We root $\T$ at node $u$, and construct another mapping $\chi': \textsf{nodes}(\T) \to 2^{\V}$ as follows. Consider an arbitrary node $u_1 \in \textsf{nodes}(\T)$. Let $\T_{u_1}$ the subtree of $\T$ rooted at $u_1$. 
        We set
        $\displaystyle{\chi'(u_1) = \chi(u_1) \cup(\y \cap  \bigcup_{u_2 \in \textsf{nodes}(\T_{u_1})} \chi(u_2))}$,
        i.e., augmenting $u_1$ with all output attributes in $\T_{u_1}$. 

        We next show $\fnfhtw(\Q) \le \rho^*(q[\y])$ by distinguishing two more cases:
        \begin{itemize}[leftmargin=*]
            \item If $u_1$ is assigned with weight $1$ by some optimal fractional edge covering of $q[\chi(u_1)]$, we have \[\rho^*\left(q[\chi(u_1)]\right) = 1 + \rho^*(q[\chi'(u_1) - \chi(u_1)]) \le  1 + \rho^*(q[\y - \chi(u)]) = \rho^*(q[\y])\] 
            where the inequality is implied by $\chi'(u_1) - \chi(u_1) \subseteq \y - \chi(u)$. Suppose not, let $A \in \left(\chi'(u_1) - \chi(u_1)\right) - \left(\y - \chi(u)\right)$ be an arbitrary attribute. Note that $A \in \chi(u)$, $A \in \chi(u_2)$ for some node $u_2 \in \nodes(\T_{u_1})$, but $A \notin \chi(u_1)$. As $u_1$ lies on the path between $t$ and $u_2$, $(\T,\chi)$ is not a valid \td for $\Q$, coming to a contradiction. 

            \item Otherwise, $u_1$ is not assigned with weight $1$ by any optimal fractional edge covering of $q[\chi(u_1)]$. We next prove $\rho^*(q[\chi(u_1) \cap \y]) = \rho^*(q[\chi(u_1)])$. Suppose not, as $\chi(u_1) \cap \y \subseteq \chi(u_1)$, we must have $\rho^*(q[\chi(u_1) \cap \y]) < \rho^*(q[\chi(u_1)])$. As both $q[\chi(u_1) \cap \y]$ and $q[\chi(u_1)]$ are acyclic, $\rho^*(q[\chi(u_1) \cap \y])$ and $\rho^*(q[\chi(u_1)])$ are integers. Meanwhile, we observe that any fractional edge covering for $q[\chi(u_1) \cap \y]$ together with assigning weight 1 to relation $e_1$ is a valid fractional edge covering for $q[\chi(u_1)]$, hence 
            $\displaystyle{\rho^*(q[\chi(u_1)]) = \rho^*(q[\chi(u_1) \cap \y]) + 1}$. Then, the optimal fractional edge covering for $q[\chi(u_1) \cap \y]$ together with assigning weight 1 to relation $e_1$ is also an optimal fractional edge covering for $q[\chi(u_1)]$, coming to a contradiction. Hence, we have: \[\rho^*(q[\chi(u_1)]) = \rho^*(q[\chi(u_1) \cap 
            \y]) \le \rho^*(q[\y]).\]
        \end{itemize}
        Applying this argument to every node in $\T$, we obtain $\fnfhtw(\Q) \le \width(\T,\chi') \le \rho^*(q[\y])$.       
    \end{proof}

    \subsection{Free-connex Submodular Width and \#Free-connex Submodular Width}
    \begin{definition}[The set of polymatroids]
    \label{defn:polymatroid}
        Given a hypergraph $q = (\V,\E$), a function $\bm h:2^{\V} \to \mathbf{D}_+$ is called a {\em polymatroid}
    if it satisfies the following properties:
    \begin{align}
        h(\bm X) + h(\bm Y) &\geq h(\bm X \cup \bm Y) + h(\bm X \cap \bm Y)
            &\forall \bm X, \bm Y \subseteq \V,  \quad
            &\text{(submodularity)}\label{eq:submod} \\
        h(\bm X) &\leq h(\bm Y)
            &\forall \bm X \subseteq \bm Y \subseteq \V\quad
            &\text{(monotonicity)}\label{eq:monotone} \\
        h(\emptyset) &= 0
        & &\text{(strictness)}\label{eq:emptyset}
    \end{align}
    The above properties are also known as {\em Shannon inequalities}.
    We use $\Gamma_{\V}$ to denote the set of all polymatroids over $\V$.
    When $\V$ is clear from the context, we drop $\V$ and write $\Gamma$.
\end{definition}

    \begin{definition}[The set of \#polymatroids]
    \label{defn:polymatroid}
        Given a hypergraph $q = (\V,\E$), a function $\bm h:2^{\V} \to \mathbf{D}_+$ is called a \#{\em polymatroid}
    if it satisfies the following properties:
    \begin{align}
        h(\bm X) + h(\bm Y) &\geq h(\bm X \cup \bm Y) + h(\bm X \cap \bm Y)
            &\forall \bm X, \bm Y \subseteq \V, \exists e\in \E, \bm X \cap \bm Y = e \quad
            &\text{(submodularity)}\label{eq:submod} \\
        h(\bm X) &\leq h(\bm Y)
            &\forall \bm X \subseteq \bm Y \subseteq \V\quad
            &\text{(monotonicity)}\label{eq:monotone} \\
        h(\emptyset) &= 0
        & &\text{(strictness)}\label{eq:emptyset}
    \end{align}
    We use $\Gamma^{\#}_{\V}$ to denote the set of all \#polymatroids over $\V$.
    When $\V$ is clear from the context, we drop $\V$ and write $\Gamma^{\#}$ instead of $\Gamma^{\#}_{\V}$.
\end{definition}

    \begin{definition}[The set of edge-dominated functions]
    \label{defn:edge-dominated}
    Given a hypergraph $q = (\V, \E)$, a function $\bm h:2^{\V} \to \mathbf{D}_+$ is called {\em edge-dominated} if it satisfies the following property: $h(\bm X) \leq 1, \forall \bm X \in \E$.
    We use $\ed_q$ to denote the set of all edge-dominated functions over $q$.
    When $q$ is clear from the context, we drop $q$ and simply write $\ed$.
    \end{definition}

\begin{definition}[Free-connex Submodular Width]
For any query $\Q$, its free-connex submodular width of $\Q$ denoted as $\fnsubw(\Q)$, is defined as:
\begin{align}
    \textsf{fn-subw}(\Q) \quad &= \quad \max_{\bm h \in \Gamma \cap \ed} \quad \min_{(\T,\chi) \in \ftd(\Q)} \quad \max_{u \in \nodes(\T)} \quad h(\chi(u))
    \label{eq:fn-subw}
    \end{align}
\end{definition}

    \begin{definition}[\#Free-connex Submodular Width]
    For any query $\Q$, its free-connex submodular width of $\Q$ denoted as $\fnsubw(\Q)$, is defined as:
    \begin{align}
    \#\textsf{fn-subw}(\Q) \quad &= \quad \max_{\bm h \in \Gamma^{\#} \cap \ed} \quad \min_{(\T,\chi) \in \ftd(\Q)} \quad \max_{u \in \nodes(\T)} \quad h(\chi(u))
    \label{eq:sharp-fn-subw}
    \end{align}
    \end{definition}

    \begin{lemma}
    \label{lem:entropy}
        Given any query $\Q = (\V, \E)$, for a subset $S \subseteq \V$ of attributes, \[\displaystyle{\rho^*(q[S])  =  \max_{\bm h \in \Gamma \cap \ed} h(S) =  \max_{\bm h \in \Gamma^{\#} \cap \ed} h(S)}.\]
    \end{lemma}

    \begin{proof}[Proof of Lemma~\ref{lem:general-ordering}]
    It suffices to show that for any query $\Q$, $\#\fnsubw(\Q) \le \fnfhtw(\Q)$. 
    For $\fnfhtw(\Q)$, we first rewrite (\ref{eq:out-width}) as below:
    \begin{align*}
            \fnfhtw(\Q) \quad = \quad \min_{(\T,\chi) \in \ftd(\Q)} \quad \width(\T,\chi)  & = \quad \min_{(\T,\chi) \in \ftd(\Q)} \quad \max_{u \in \nodes(\T)} \quad \rho^*(\chi(u)) \\
            & = \quad \min_{(\T,\chi) \in \ftd(\Q)} \quad \max_{u \in \nodes(\T)} \quad \max_{\bm h \in \Gamma^{\#} \cap \ed}  \quad h(\chi(u)) \\
            & = \quad \min_{(\T,\chi) \in \ftd(\Q)} \quad \max_{\bm h \in \Gamma^{\#} \cap \ed}  \quad \max_{u \in \nodes(\T)} \quad h(\chi(u))
        \end{align*}

        By comparing the above and (\ref{eq:sharp-fn-subw}), it immediately follows from the min-max inequality that $\#\fnsubw(\Q) \le \fnfhtw(\Q)$.
    \end{proof}

    \subsection{Proof of Lemma~\ref{lem:equivalence}}
        Implied by Lemma~\ref{lem:general-ordering}, it suffices to show that for any acyclic query $\Q$, $\fnsubw(\Q) \ge \fnfhtw(\Q)$ holds. Recall the procedural definition of $\fnfhtw(\Q)$ in Corollary~\ref{cor:recursive}. We prove it via two steps:

        \paragraph{Step 1: $\Q$ is $\exists$-connected} In this case, we will identify a polymatroid $\bm h \in \Gamma \cap \ed$ such that for any \td $(\T,\chi) \in \ftd(\Q)$, there always exists a node $u \in \nodes(\T)$ with $h(\chi(u)) = \rho^*(q[\y])$. Note that $\fnfhtw(\Q) = \rho^*(q[\y])$. From Lemma~\ref{lem:independent-S}, let $S \subseteq \y$ be a set of $\fnfhtw(\Q)$ output attributes such that no pairs of them appear in the same relation from $\E$. We identify a modular function $h$ as follows. Let each attribute $A \in S$ represent an independent random variable with $h(A) = 1$. Let each remaining attribute $B \in \V -S$ represent a constant with $h(B) = 0$. It can be easily checked that $h \in \Gamma \cap \ed$. Consider an arbitrary \td $(\T,\chi) \in \ftd(\Q)$. From Lemma~\ref{lem:root}, there must exist a node $u \in \nodes(\T)$ such that $\chi(u) =\y$. Hence, $h(\chi(u)) = \sum_{A \in S}h(A) = |S| = \fnfhtw(\Q) = \rho^*(q[\y])$.

        \paragraph{Step 2: $\Q$ is $\exists$-disconnected} Let $\Q_1,\Q_2,\cdots, \Q_h$ be the $\exists$-connected subqueries of $\Q$. Wlog, assume $\fnfhtw(\Q)= \fnfhtw(\Q_1)$. 
        Let $\Q_i = (\V_i, \E_i,\y_i)$ where $\V_i = \bigcup_{e \in \E_i} e$ and $\y_i = \V_i \cap y$. 
  
        Similarly, let $S \subseteq \y_1$ be a set of $\fnfhtw(\Q_1)$ output attributes such that no pair of them appears in the same relation from $\E_1$. We next argue that no pair appears in the same relation from $\E$. Suppose not, assume $A, A' \in S$ are two output attributes that appear together in some relation $e \in \E - \E_1$. In $\Q_1$, let $A_0(=A),A_1,A_2,\cdots, A_\ell,A_{\ell+1} (=A')$ be a sequence of attributes such that (1) $A_1,A_2,\cdots, A_\ell \in \V -\y$; (2) there exists $\ell+1$ relations $e_0,e_1,\cdots, e_{\ell} \in \E_1$ such that for each $i \in \{0,1,\cdots, \ell\}$, we have $A_i,A_{i+1} \in e_i$; (3) for every pair of non-consecutive attributes $A_i,A_j$ for some $i,j$ with $|j-i| > 1$, there exists no relation $e' \in \E_1$ containing both $A_i,A_j$. 
        Note that for any other relation $e \in \E- \E_1$, $e \cap \{A_1,A_2,\cdots, A_\ell\} = \emptyset$, which is implied by the definition of $\exists$-connectivity. Hence, (3) is equivalent to for every pair of non-consecutive attributes $A_i,A_j$ for some $ I,j$ with $|j-i| > 1$, there exists no relation $e' \in \E$ containing both $A_i,A_j$. If $\ell \ge 1$, we have identified a cycle $A_0(=A),A_1,A_2,\cdots, A_\ell,A_{\ell+1} (=A')$, contradicting the fact that $\Q$ is acyclic. Hence, $\ell < 1$, which means $A_1,A_2,\cdots,A_\ell$ do not exist; in other words, $A,A'$ appear in some common relation from $\E_1$.
        As $\Q$ is acyclic, there must exist a relation $e' \in \E$ such that $A, A_1, A_2, \cdots, A_\ell, A' \in e'$; otherwise, we have identified a cycle, contradicting the fact that $\Q$ is not acyclic. As $A_1,A_2,\cdots, A_\ell \in e'$, $e' \in \E_1$. This comes to a contradiction that $A,A'$ does not appear in the same relation of $\E_1$. Again, let each attribute $A \in S$ represent an independent random variable with $h(A) = 1$. Let each remaining attribute $B \in \V -S$ represent a constant with $h(B) = 0$. It can be easily checked that $h \in \Gamma \cap \ed$. We claim that a node $u \in \nodes(\T)$ must exist with $\y_1 \subseteq \chi(u)$. Consider a \td $(\T,\chi')$ with $\chi'(v)= \chi(v) - (\y -\y_1)$ for each node $v \in \nodes(\T)$. Note that $(\T,\chi')$ is also a valid free-connex \td for $\Q_1$; hence there must exist some node $u \in \nodes(\T)$ with $\chi'(u) = \y_1$. Correspondingly, $\y_1 \subseteq \chi(u)$. Hence, $h(\chi(u)) = \sum_{A \in S}h(A) = |S| = \fnfhtw(\Q)$.

    \section{Yannakakis algorithm Revisited}
    \label{appendix:yannakakis}
    \subsection{Missing pseudocode of the Yannakakis Algorithm}

    We give the pseudocode of the Yannakakis Algorithm in Algorithm~\ref{alg:yannakakis}, which takes as input an acyclic join-aggregate query $\Q$ and an instance $\R$.
      \begin{algorithm}[h]
    \caption{$\textsc{Yannakakis}(\Q =(\V,\E,\y), \R)$~\cite{yannakakis1981algorithms,joglekar16:_ajar}}
    \label{alg:yannakakis}
    Let $(\T, \chi)$ be a width-1 \td of $\Q$ rooted at node $r$\;
    \ForEach{node $u \in \nodes(\T)$}{
        \lIf{$\exists e \in \E$ s.t. $\chi(u) = e$}{$R_u \gets R_e$}
        \lElse{$R_u \gets \bigcap_{e \in \E} \pi_{\chi(u)} R_e$ where each tuple $t \in R_u$ has its annotation as $1$}}
    \While{visit nodes a bottom-up way (excluding the root $r$)}{
        \ForEach{node $u$ visited}{
            $u' \gets$ the parent of $u$\; 
            $R_{u'} \gets R_{u'} \ltimes R_u$\;
        }
    }
    \While{visit nodes a top-down way (excluding the root $r$)}{
        \ForEach{node $u$ visited}{
            $u' \gets$ the parent of $u$\; 
            $R_{u} \gets R_{u} \ltimes R_{u'}$\;
        }
    }
     \While{visit nodes a bottom-up way (excluding the root $r$)}{
        \ForEach{node $u$ visited}{
            $u' \gets$ the parent of $e$\;
            $R_u \gets \oplus_{ \chi(u) - \chi(u')-\y} R_u$\;
            $R_{u'} \gets R_u \Join R_{u'}$\;
        }
    }
    \Return $\oplus_{\chi(r) -\y} R_r$ for the root $r$\;
    \end{algorithm}

    \subsection{Discussion on \cite{khamis17:_what}}

      As mentioned, the algorithm in \cite{khamis17:_what} also has time complexity $O\left(N^{\fnfhtw(\Q)} + \OUT\right)$. For any acyclic query $\Q = (\V,\E,\y)$ with $q = (\V,\E)$, $\fnfhtw(\Q) \le \rho^*(q[\y])$. Hence, we can show that $N^{\fnfhtw(\Q)} + \OUT = O(N^{\rho^*(q[\y])})$. Therefore, \cite{khamis17:_what} is worst-case optimal for acyclic queries.

     On the other hand, we note that $\displaystyle{N \cdot \OUT^{1-\frac{1}{\fnfhtw(\Q)}} + \OUT = O\left(N^{\fnfhtw(\Q)} + \OUT\right)}$, since $N \cdot \OUT^{1- \frac{1}{\fnfhtw(\Q)}} \le N^{\fnfhtw}(\Q)$ always holds when $\OUT< N \cdot \OUT^{1-\frac{1}{\fnfhtw(\Q)}}$. Moreover, when $\OUT < N^{\fnfhtw(\Q)}$, this algorithm is $\left(\frac{N^{\fnfhtw(\Q)}}{\OUT}\right)^{1 - \frac{1}{\fnfhtw(\Q)}}$-factor worse than our algorithm.

    \subsection{Worst-case Optimality of  Yannakakis Algorithm on Acyclic Queries}      
    
    Next, we will give a new analysis of the Yannakakis algorithm by a more fine-grained inspection of its execution. Similar to Algorithm~\ref{alg:meta}, we first remove all dangling tuples, decompose the query into a set of $\exists$-connected subqueries, compute their query results separately, and finally combine their results via joins. Below, we focus on $\exists$-connected acyclic queries.
    
    In the proof of Lemma~\ref{lem:out-width-connected}, we have identified a width-1 \td $(\T,\chi)$ such that its corresponding free-connex \td $(\T',\chi')$ defined by the Yannakakis algorithm has $\width(\T,\chi') = \fnfhtw(\Q)$. Recall that for each node $u \in \nodes(\T)$, $\chi'(u)$ is the set of attributes on which the intermediate join results will be materialized by the Yannakakis algorithm. Hence, the intermediate join size can be bounded by $O\left(N^{\fnfhtw(\Q)}\right)$. We obtain:

    \begin{theorem}
    \label{the:yannakakis-worst-case}
      For any acyclic query $\Q =(\V,\E,\y)$ and an instance $\R$ of input size $N$ and output size $\OUT$, the Yannakakis algorithm can compute $\Q(\R)$ within $O\left(N^{\fnfhtw(\Q)} + \OUT\right)$ time.
    \end{theorem}

    Similar to the above, the Yannakakis algorithm is also worst-case optimal for acyclic queries.

    \subsection{a-hierarchical Query}
    \label{appendix:hierarchical}

    \begin{definition}[Hierarchical Query~\cite{suciu2011probabilistic}] 
    A query $\Q=(\V,\E,\y)$ is   hierarchical if for any pair of attributes $A,B \in \V$, either $\E_{A} \subseteq \E_{B}$, or $\E_{B} \subseteq \E_{A}$, or $\E_{A} \cap \E_{B} = \emptyset$.
   \end{definition}
    
    A hierarchical query must be acyclic, and a free-connex query must also be acyclic. However, there is no containment relationship between free-connex and hierarchical queries.

    \begin{definition}[a-hierarchical Query]
    \label{def:a-hierarchical}
        A query $\Q$ is a-hierarchical if $\Q$ is acyclic, and every $\exists$-connected subquery $\Q'_1, \Q'_2, \cdots, \Q'_\ell$ of $\Q'$ is hierarchical, where $\Q'$ is the cleansed version of $\Q$.
    \end{definition}
    
    \begin{theorem}
        \label{the:hierarchical}
        For any a-hierarchical query $\Q$ and an instance $\R$ of input size $N$ and output size $\OUT$, the Yannakakis algorithm computes $\Q(\R)$ in $O\left(N \cdot \OUT^{1-\frac{1}{\fnfhtw(\Q)}} + \OUT\right)$ time. 
    \end{theorem}

          \begin{wrapfigure}{r}{0.7\linewidth}
        \centering
        \vspace{-1em}
            \includegraphics[scale=0.9]{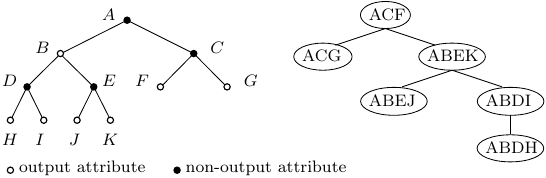}
            \caption{An illustration of the attribute tree (left) and width-1 \td  (right) of a cleansed, $\exists$-connected, and hierarchical join-aggregate query $\Q = \oplus_{A,C,D,E} R_1(A,B,D,H) \Join R_2(A,B,D,I) \Join R_3(A,B,E,J) \Join R_4(A,B,E,K) \Join R_5(A,C,F) \Join R_6(A,C,G)$.}
        \label{fig:atttribute-tree}
        \vspace{-1em}
    \end{wrapfigure}
    We also resort to the same outline as Algorithm~\ref{alg:meta}, except that we invoke the Yannakakis algorithm at lines 4-5. It suffices to assume that the input join-query $\Q = (\V, \E,\y)$ is acyclic, cleansed, and existentially connected.
    By definition, $\Q$ is hierarchical. 
    Moreover, $\Q = (\V,\E,\y)$ has an attribute tree $\H$, such that (\romannumeral 1) there is a one-to-one correspondence between attributes in $\V$ and nodes in $\H$; (\romannumeral 2) each relation corresponds to a leaf-to-root path; (\romannumeral 3) every leaf node (that corresponds to a unique attribute) must be an output attribute. 
    Note that $|\E| = \fnfhtw$. In $\H$, let $\H_x$ be the subtree of $\H$ rooted at attribute $x \in \V$. Let $\Q_x$ be the sub-query derived by relations that contain attribute $x$, i.e., $\Q_x = (\V_x, \E_x, \y_x)$, where $\V_x = \bigcup_{e \in \E_x} e$ and $\y_x = \y \cap \V_x$. 
    Let $\textsf{path}(x_1,x_2)$ denote the set of attributes lying on the path between $x_1$ and $x_2$ in $\H$. 
    
    We build a width-1 \td $(\T,\chi)$ for $\Q$ as follows. 
    Consider the children of the root attribute in $\H$.
    Let $\{\E_1,\E_2,\cdots, \E_j\}$ be a partition of $\E$ such that all relations in $\E_i$ shares one common child attribute. We build a width-1 \td  $(\T_i,\chi_i)$ for each group $\E_i$, and then add $\T_i$ for every $i <j$ as the last $(j-1)$ child nodes of the root of $\T_j$. See an example in Figure~\ref{fig:atttribute-tree}. We apply the Yannakakis algorithm along such a width-1 \td~  by traversing nodes in a post order. 
    
    Below, we aim to bound the number of intermediate join results materialized by the Yannakakis algorithm.  Let $q = (\V,\E)$ be the corresponding join query of $\Q$.
    For simplicity, we define
    \[\lambda_{q,\y, \R}(x) = \left|\pi_{\textsf{path}(x,r) \cup \y_x} q(\R)\right|\] 
    for attribute $x\in \V$.
    The number of intermediate join result that materialized is exactly $\displaystyle{O\left(\max_{x \in \H} \lambda_{\R}(x)\right)}$. Let $\Re_{\Q}(N,\OUT)$ denote the class of all input instances of input size $N$ and output size $\OUT$ for $\Q$.
    We can further rewrite the size bound above as:
    \begin{align*}
        \max_{\R \in \Re_{\Q}(N,\OUT)}  \max_{x \in \H} \lambda_{\R}(x) 
        \le \max_{x \in \H} 
        \max_{\R \in \Re_{\Q}(N,\OUT)} \lambda_{\R}(x)
    \end{align*}
    Then, it suffices to prove for an arbitrary 
    attribute $x \in \H$:
    $\displaystyle{\max_{\R \in \Re_{\Q}(N,\OUT)} \lambda_{\R}(x) \le 
    N \cdot \OUT^{1-\frac{1}{|\E_x|}}}$. 
    As $|\E_x| \le \fnfhtw$,
    we come to the desired result. Let's take a closer look at the sub-query $\lambda_{\R}(x)$ derived, which is captured by the class of {\em generalized star query} below. Implied by Lemma~\ref{lem:generalized-star-intermediate}, we complete the whole proof.

    \begin{definition}[Generalized Star]
    \label{def:generalized-star}
        A cleansed, $\exists$-connected and hierarchical query $\Q = (\V,\E,\y)$ is a {\em generalized star} if $\V -\y \subseteq \bigcap_{e \in \E} e$. 
    \end{definition}   

    \begin{lemma}
    \label{lem:generalized-star-intermediate}
        For a generalized star query $\Q =(\V, \E,\y)$ of $k$ relations, given any parameter $1\le N$ and $\OUT \le N^k$, we have $\displaystyle{\max_{\R \in \Re_{\Q}(N,\OUT)} \left|\Join_{e \in \E} R_e\right| = O\left(N \cdot \OUT^{1-\frac{1}{k}}\right)}$.
    \end{lemma}

    \begin{proof}
        We distinguish two more cases on $\Q$. If $\bigcap_{e \in \E} e \subseteq \y$. In this case, $\left|\Join_{e' \in \E} R_{e'}\right| = \OUT$ holds for an arbitrary instance $\R \in \Re_{\Q}(N,\OUT)$. As $\OUT \le \prod_{e \in \E}|R_e|$, this result automatically holds. In the remaining, we focus on the case when $\bigcap_{e \in \E} e -\y \neq \emptyset$. Let $\mathbf{z} = \bigcap_{e \in \E} e \cap \y$. Our proof consists of two steps:
        
       \paragraph{Step 1} Consider a derived sub-query $\Q' = (\V', \E', \y')$ where $\V' = \V - \mathbf{z}$, $\E' = \{e \cap \V': e \in \E\}$ and $\y' = \y - \mathbf{z}$. Note that $|\E'| = |\E|$. We will prove
            \[\displaystyle{\max_{\R \in \Re_{\Q'}( N,\OUT)} \left|\Join_{e \in \E} R_e\right| = O\left(N \cdot \OUT^{1-\frac{1}{|\E'|}}\right)}\]
        with a similar argument made in \cite{amossen2009faster}.
        Let $\bigcap_{e \in \E'} e = \{B\}$ be the unique non-output attribute appearing in all relations from $\Q'$.
        Suppose $\dom(B) = \{b_1,b_2,\cdots, b_\ell\}$. We introduce a variable $\triangle^e_i$ for each value $b_i$ to denote the number of input tuples that display $b_i$ in attribute $B$ in relation $R_e$. 
        The largest number of full join results can be produced 
        is captured as follows:
        \begin{align*}
            \textrm{max.} \ \ \ \sum_{i \in [\ell]} \prod_{e \in \E'} \triangle^{e}_{i} \ \ \ \textrm{subject to.} \ \ \ & \sum_{e\in \E'}\sum_{i \in [\ell]} \triangle^e_i \le N, \ \ \ \prod_{e \in \E'}\triangle^e_i \le \OUT, \ \ \forall i \in [\ell]
        \end{align*}
        The optimal solution $\Theta\left(N \cdot \OUT^{1- \frac{1}{|\E'|}}\right)$ is achieved when $\triangle^e_i = \OUT^{\frac{1}{|\E'|}}$ for any $i \in [\ell]$ and $e\in \E'$, and $\ell = \frac{N}{|\E'| \cdot \OUT^{\frac{1}{|\E'|}}}$.
        
        \paragraph{Step 2} Consider an arbitrary instance $\R$ of $\Q$. Let $z_1, z_2,\cdots, z_\ell$ be the values in the effective domain of $\mathbf{z}$ in $\R$. Let $\displaystyle{N_i = \sum_{j \in [\ell]} \left|\sigma_{\mathbf{z} = z_i} R_j\right|}$ and $\displaystyle{\OUT_i = \prod_{j \in [\ell]} \left|\sigma_{\mathbf{z} = z_i} \Q(\R)\right|}$. Implied by {\bf Step 1}, the largest number of full join results can be produced is at most $O\left(N_i \cdot \OUT_i^{1-\frac{1}{|\E|}}\right)$. Given the following two constraints $\displaystyle{\sum_{i \in [\ell]} N_i = N, \textrm{ and } \sum_{i \in [\ell]} \OUT_i = \OUT}$, the largest number of full join result produced can be bounded by
        $\sum_{i \in [\ell]} N_i \cdot \OUT_i^{1-\frac{1}{|\E|}} \le  \sum_{i \in [\ell]} N_i \cdot \OUT^{1-\frac{1}{k}} \le N \cdot \OUT^{1-\frac{1}{|\E|}}.
        $
        As $|\E| = k$, this is exactly $O\left(N \cdot \OUT^{1-\frac{1}{k}}\right)$.

        \smallskip
        Combining these two steps, we complete the whole proof.
   \end{proof}
    
    \subsection{Non-a-hierarchical Query}
    \label{appendix:non-hierarchical}
\begin{wrapfigure}{r}{0.36\linewidth}
         \centering
         \vspace{-0.5em}
        \includegraphics[scale=0.75]{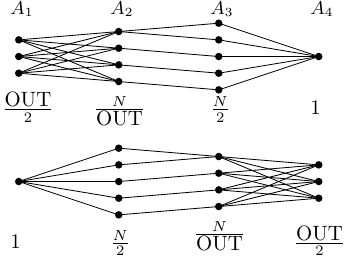}
        \vspace{-1em}
        \caption{An illustration of hard instance for Yannakakis algorithm on line query with $k=3$~\cite{hu2024fast}.}
        \label{fig:line}
    \end{wrapfigure}

    As shown in \cite{hu2024fast}, Yannakakis algorithm indeed incurs $\Theta(N\cdot \OUT)$ time for line queries. We first revisit the hard instance constructed for line-3 query as shown in Figure~\ref{fig:line}. Consider an arbitrary acyclic but non-a-hierarchical join-aggregate query $\Q$. Let $\Q'$ be its cleansed version. Let $\Q_1,\Q_2,\cdots,\Q_h$ be the connected sub-queries in $G^\exists_{\Q'}$. As $\Q$ is acyclic, every sub-query $\Q_i$ is also acyclic. As $\Q$ is non-a-hierarchical, at least one sub-query is not hierarchical, say $\Q_1$. There must exist a pair of attributes $x,x'$ and three relations $e,e',e''$ such that $x \in e \cap e' - e''$ and $x' \in e' \cap e'' - e$. Let $(\T,\chi)$ be a width-1 \td  of $\Q_1$. As $\Q_1$ is cleansed, every leaf node must contain some unique attribute. 

    It is always feasible to find a pair of leaf nodes $e_1,e_n$ in $\T$ such that $e,e'' \in \textsf{path}(e_1,e_n)$, $e \in \textsf{path}(e_1,e'')$ and $e'' \in \textsf{path}(e',e_n)$. It could be possible that $e = e_1$ or $e''= e_n$. Let $A_1, A_{n+1}$ be an arbitrary unique output attribute in $e_1,e_n$ separately. As $\Q_1$ is $\exists$-connected, there must exists a subset $S$ of $m$ relations in $\textsf{path}(e_1,e_n)$ (including $e_1$ and $e_n$) and a subset of non-output attributes $B_1, B_2,\cdots, B_m$ such that $B_1 \in e_1$, $B_n \in e_n$, $B_i, B_{i+1} \in e_i$ for some relation $e_i \in \textsf{path}(e_1,e_n)$, and there exists no relation $e'$ such that $B_i, B_j \in e'$ for any $|j -i| > 1$. Otherwise, $\Q_1$ is not $\exists$-connected. 

    Give an instance $\R_\textsf{line}$ of $\line$ with $k=3$, we construct an instance $\R$ for $\Q$ as follows. We set $\dom(x) = \{*\}$ for every attribute $x \in \V - \{C_1,C_2,B_1,B_2,\cdots,B_m\}$. We use $C_1$ to simulate $A_1$, use $C_2$ to simulate $A_4$, $B_1$ to simulate $A_2$ and all remaining attributes $B_2,B_3,\cdots, B_m$ to simulate $A_3$. For any relation $e$, such that $\left|e \cap \{B_2,B_3,\cdots, B_m\}\right| > 1$, the projection of $R_e$ onto $\{B_2,B_3,\cdots, B_m\}$ should be a one-to-one mapping. The argument for line-3 query can be applied for $\Q$ similarly, i.e., any query plan of the Yannakakis algorithm requires $\Theta(N \cdot \OUT)$ time.
    
    \section{Cyclic Queries}
    \label{appendix:cyclic}
    We can obtain output-sensitive algorithms for cyclic queries by replacing the Yannakakis algorithm with our new output-optimal algorithm. Given a query $\Q = (\V,\E,\y)$, a \td $(\T, \chi)$ also defines another query $\Q_\T = (\V, \{\chi(u): u \in \nodes(\T)\}, \y)$. Each instance $\R$ for $\Q$ defines a derived instance $\R_\T$ as follows. For each $e \in \E$, we assign it to one specific node $u \in \nodes(\T)$ such that $e \subseteq \chi(u)$. For each node $u \in \nodes(\T)$, we get a relation $R_u:= \Join_{e \in \E}\pi_{\chi(u) \cap e} R_e$ as the result of the corresponding sub-query $q[\chi(u)]$. For each $e \in \E$ with $\chi(u) \cap e \neq \emptyset$, we distinguish the following two cases: if $e$ is not assigned to $t$, each tuple in the sub-relation has its annotation as $1$; otherwise, each tuple in $\pi_{\chi(u) \cap e} R_e$ has its annotation as the same in $R_e$. 
    
    \begin{theorem}
    \label{the:cyclic-1}
        For an arbitrary query $\Q = (\V,\E,\y)$, and an instance $\R$ of input size $N$ and output size $\OUT$, $\Q(\R)$ can be computed in $\displaystyle{O\left(\min_{(\T, \chi)} N^{\width(\T,\chi)} \cdot \OUT^{1-\frac{1}{\fnfhtw(\Q_\T)}} + \OUT\right)}$
        time, where $(\T,\chi)$ is over all possible \tds~ of $\Q$.
    \end{theorem}

    As the query size is a constant, we can afford to find the optimal \td $(\T,\chi)$ in $O(1)$ time. People have also exploited the power of hybrid \tds~ to speed up query evaluation in the literature. Similar to the idea of~\cite{khamis17:_what}, we first define the notion of {\em \textsf{TD}-coverage} as a set of \tds $(\T_1,\chi_1), (\T_2, \chi_2), \cdots, (\T_h, \chi_h)$ such that $\Q(\R) = \bigoplus_{i \in [h]} \Q_{\T_i}(\R_{\T_i})$. Similarly, we can apply our acyclic algorithm to each $(\T_i, \chi_i)$ and aggregate the query results over all \tds.

    \begin{theorem}
    \label{the:cyclic-1}
        For an arbitrary query $\Q = (\V,\E,\y)$, and any instance $\R$ of input size $N$ and output size $\OUT$, $\Q(\R)$ can be computed in \[\displaystyle{O\left(\min_{\{(\T_1,\chi_1), (\T_2,\chi_2), \cdots, (\T_h,\chi_h)\}} \max_{i \in [h]}N^{\width(\T_i,\chi_i)} \cdot \OUT^{1-\frac{1}{\fnfhtw(\Q_{\T_i})}} + \OUT\right)}\]
        time, where $\{(\T_1,\chi_1),(\T_2,\chi_2),\cdots, (\T_h,\chi_h)\}$ is over all possible \textsf{TD}-coverages of $\Q$.
    \end{theorem}
    If we only need to consider the parameter $\width(\T_i,\chi_i)$, the minimum \td decompositions can be identified by \cite{khamis17:_what}, and the cost of this approach is captured by the \#submodular width! However, we need to consider both $\width(\T_i,\chi_i)$ and $\fnfhtw(\Q_{\T_i})$ to minimize the formula above. This question is interesting but very challenging, which we leave as an open question.

    \section{Missing Proofs in Section~\ref{sec:acyclic}}
    \label{appendix:acyclic}
    \begin{lemma}
    \label{lem:assignment}
        There exists a subset $\E' \subseteq \E$ of $\fnfhtw$ relations such that $\y \subseteq \bigcup_{e \in \E'} e$.
    \end{lemma} 

    \begin{proof}[Proof of Lemma~\ref{lem:assignment}]
        As $\Q$ is acyclic, $q[\y]$ is also acyclic. As shown in~\cite{hu2021cover}, every acyclic query has an optimal fractional edge covering $\rho^*$ that is also integral, i.e., $\rho^*(e) = 1$ or $\rho^*(e) = 0$ for any relation $e \in \E$. Let $\E' \subseteq\E$ be the set of relations for which $\rho^*(e) = 1$ holds for every $e \in \E'$. For every attribute $A \in \y$, there must exist a relation $e \in \E'$ such that $A \in e$. Implied by the definition of out-width, $|\E'| = \fnfhtw$. 
    \end{proof}

    \begin{proof}[Proof of Lemma~\ref{lem:leaves}]
         We start with an arbitrary width-1 \td $(\T,\chi)$ for a separated acyclic query $\Q$. Let $\E_\bullet = \{e \in \E: e \cap \y \neq \emptyset\}$ be the set of relations containing some (unique) output attributes. If there exists some relation $e \in \E_\bullet$ that is not a leaf node of $\T$, we perform the following transform procedure.  As $\Q$ is separated, there exists a relation $e' \in \E - \{e\}$ such that $e-\y \subset e'$. Hence, we move every child node $e''$ of $e$ together with the subtree rooted at $e''$ as a new child node of $e'$. It can be checked that the resulting tree is still a valid width-1 \td\ for $\Q$. Now, $e$ becomes a leaf node. 

         After this step, we can argue that every leaf node of $\T$ belongs to $\E_\bullet$. Suppose not, there exists a leaf node $e'$ of $\T$ but $e' \notin \E_\bullet$. Let $e''$ be the (unique) neighbor node of $e'$. There must be the case that all unique attributes in $e'$ are non-output attributes, and all joint attributes in $e'$ are also contained by $e''$. Hence, $e'$ should have been removed by the cleansed procedure, contradicting that $\Q$ is cleansed. Together, $\E_\bullet$ is exactly the set of leaf nodes of $\T$. 
    \end{proof} 
    
    \begin{proof}[Proof of Lemma~\ref{lem:large-reverse-limited}]
    Consider an arbitrary tuple $t \in \dom(e_1 \cap e_2)$. As $(e_1,e_2)$ is large, $t$ can be joined with at least $\OUT^{\phi_{e_1,e_2}}$ query result of $\Q_{e_1,e_2}$. After removing dangling tuples, every tuple $t' \in \pi_{\y} \Q_{e_2,e_1}$ can appear together with at least $\OUT^{\phi_{e_1,e_2}}$ tuples from $\pi_{\y -e_1} \Q_{e_1,e_2}$ in the final query result. This way, $|\bigoplus_{\V-\y} \Q_{e_2,e_1}|\le \OUT^{1- \phi_{e_1,e_2}} = \OUT^{\phi_{e_2,e_1}}$. Hence, $(e_2,e_1)$ is limited.
    \end{proof}

    \begin{proof}[Proof of Lemma~\ref{lem:limited-imply-limited}]
    For $\Q_{e_1,e_2}$, we observe the following:  
    \begin{align*}
    \left|\bigoplus_{\V-\y}  \Q_{e_1,e_2}\right| & \le \prod_{e_3 \in \mathcal{N}_{e_1} - \{e_2\}} \left|\bigoplus_{\V-\y} \Q_{e_3, e_1}\right| \le \prod_{e_3 \in \mathcal{N}_{e_1} - \{e_2\}} \OUT^{\phi_{e_3,e_1}} \le  \OUT^{\phi_{e_1,e_2}}
    \end{align*}
    where the first inequality follows that $\displaystyle{\bigcup_{e \in \T_{e_1,e_2}}(e \cap \y) = \bigcup_{e_3 \in \mathcal{N}_{e_1} - \{e_2\}} \bigcup_{e \in \T_{e_1,e_2}} (e\cap \y)}$ and the second inequality follows that $\displaystyle{\sum_{e_3 \in \mathcal{N}_{e_1} - \{e_2\}} \phi(e_3,e_1) = \phi(e_1,e_2)}$. By definition, $(e_1,e_2)$ must be limited. 
    \end{proof}

     \begin{lemma}[Not-All-Large]
        \label{lem:not-all-large}
        Consider a connected subtree $\T_1$ of $\T$ that does not contain any leaf node of $\T$, but removing $\T_1$ turns $\T$ into a set of disconnected sub-trees. For any instance $\R'$ with non-empty query result, there must exist a pair of nodes $e_1 \in \T_1, e_2 \in \T - \T_1$ such that $(e_1,e_2)$ is an edge in $\T$ but not large.
    \end{lemma}
    
    \begin{proof}[Proof of Lemma~\ref{lem:not-all-large}]
    Let $\V_1$ be the set of nodes in $\T_1$ that are incident to some nodes in $\T_2$. Let $\V_2$ be the set of nodes in $\T - \T_1$ that are incident to some nodes in $\T_1$. Note that there is a one-to-one correspondence between $\V_1$ and $\V_2$. Let $e_2 \in \V_2$ be the corresponding node for $e_1 \in \V_1$. By contradiction, suppose every edge $(e_1,e_2)$ is large for every $e_1 \in \V_1$. Then, edge $(e_2,e_1)$ must be limited. Consider an arbitrary node $e^*_1 \in \V_1$ with its corresponding node $e^*$. We note that for every tuple in $t \in \dom(e^*_1 \cap e^*_2)$,
    \begin{align*}
        \left|\sigma_{e^*_1 \cap e^*_2 =t} \Q_{e^*_1, e^*_2}(\R)\right| \le \prod_{e_1 \in \V_1} \left|\pi_{\y} \Q_{e_2, e_1}(\R)\right| \le \prod_{e_1 \in \V_1 - \{e^*_1\}} \Tilde{\OUT}^{\phi_{e_2,e_1}} \le \Tilde{\OUT}^{\phi_{e^*_1,e^*_2}}
    \end{align*}
    where the last inequality is implied by the fact that $\displaystyle{\sum_{e_1 \in \V_1 - \{e^*_1\}}\left|\E_\bullet \cap \T_{e_2,e_1}\right|= \left|\E_\bullet \cap \T_{e^*_1, e^*_2}\right|}$. Hence, edge $(e^*_1, e^*_2)$ is not large, coming to a contradiction.    
    \end{proof}

    \begin{algorithm}[t]
        \caption{\textsc{IdentifyLeaf}$(\Q, \R', \T')$}
        \label{alg:identify}
        \SetKwInOut{Input}{Input}
        \SetKwInOut{Output}{Output}

        \Input{A separated acyclic query $\Q = (\V,\E,\y)$, an instance $\R'$ and a (partially) labeled separated width-1 \td  $\T'$, in which no more edges can be labeled by Algorithm~\ref{alg:partition};}
        \Output{A leaf node as described in Lemma~\ref{lem:base};}
        $\E_\bullet \gets \{e \in \E: e\cap \E \neq \emptyset\}$\;
        \While{$\T'$ is not a single node}{
            Root $\T'$ at an arbitrary node $r \in \T' \cap \E_\bullet$\;
            \lForEach{$e \in \T'$}{$p_e \gets$ the parent node of $e$}
            \If{$\exists$ $e \in \T'$ such that edge $(e,p_e)$ is large but edge $(e',p_{e'})$ is small for every other node $e'$ in the subtree rooted at $e$}{
                $\T' \gets \T' - \T'_{e, p_e}$\;
            }
            \lElse{\Return $r$}
        }
        \end{algorithm}

     To show the correctness of Algorithm~\ref{alg:partition}, we first need to prove:

    \begin{lemma}
    \label{lem:partition-correctness}
        In Algorithm~\ref{alg:partition}, for any $(\T',\R') \in \mathcal{P}$, either line 8, or line 10, or line 13 can be applied.
    \end{lemma}

    \begin{proof}[Proof of Lemma~\ref{lem:partition-correctness}]
        Consider an arbitrary instance $\R'$ with a partially labeled separated width-1 \td  $(\T',\chi')$. By contradiction, we assume that neither line 8 nor 10 of Algorithm~\ref{alg:partition} can be applied. We next show how to identify a leaf node $e$ of $\T'$ such that the edge $(*,e)$ is small; hence, $\R'$ already meets the optimal condition, coming to a contradiction. As shown in Algorithm~\ref{alg:identify}, the high-level idea is to conceptually remove nodes in $\T'$, until a single node is left, which is exactly the leaf node as desired. We root $\T'$ at an arbitrary node $r \in \E_\bullet \cap \T'$. Let $p_e$ be the parent of $e$ in this rooted tree. If there exists a node $e$ such that the edge $(e,p_e)$ is large, but the edge $(e',p_{e'})$ is small for every node $e'$ residing in the subtree rooted at $e$, we simply remove the whole subtree $\T'_{e,p_e}$ and continue. Otherwise, every edge $(e,p_e)$ is small, including the edge $(*,r)$ incident to $r$, hence $r$ is returned. 
        
       It remains to show that an arbitrary node $r \in \T' \cap \E_\bullet$ can always be found at line 3. We note that $\T'$ is always connected in the execution process. Moreover, $\T' \cap \E_\bullet \neq \emptyset$. To show this, we need Lemma~\ref{lem:not-all-large}. By contradiction, assume $\T' \cap \E_\bullet = \emptyset$. Let $\T'_1$ be the subtree(s) removed from the initial $\T'$. As $\T'$ is connected and $\T' \cap \E_\bullet = \emptyset$, $\T'_1$ must be disconnected. Moreover, for each pair of nodes $e_1 \in \T'$, $e_2 \in \T'_1$, the subtree $\T'_{e_2,e_1}$ has been removed due to the fact that edge $(e_1,e_2)$ is large. This way, Lemma~\ref{lem:not-all-large} is violated on $\T'$, coming to a contradiction. Hence, $\T' \cap \E_\bullet \neq \emptyset$ always holds in the execution process. 
       This means that it is always feasible to root $\T'$ at some node in $\T' \cap \E_\bullet$, and further shrink $\T'$. As $\T'$ has a limited size, it will finally end with the case that every edge $(e,p_e)$ is small for any $e \in \T'$.
    \end{proof}

    \begin{lemma}
        \label{lem:while-lopp}
         Algorithm~\ref{alg:partition} will terminate after running the while-loop at most $O(2^{|\E|})$ iterations. 
    \end{lemma}
    
    \begin{proof}[Proof of Lemma~\ref{lem:while-lopp}]
        As each edge will be labeled once, and labeling each edge can lead to at most $2$ sub-instances, the total number of sub-instances in $\mathcal{P}$ is $O(2^{|\E|})$, where the number of edges in $\T$ is bounded by the number of relations in $\Q$. Each iteration removes at least one sub-instance from $\mathcal{P}$. Moreover, when a sub-instance is removed, it won't be added back to $\mathcal{P}$. Hence, $\mathcal{P}$ will become $\emptyset$ after at most $O(2^{|\E|})$ iterations.
    \end{proof}

      Next, we analyze the runtime of Algorithm~\ref{alg:partition}. Let's first focus on line 14. For any node $e_3 \in \mathcal{N}_{e_1} - \{e_2\}$, as edge $(e_3,e_1)$ is small, $\Q_{e_3, e_1}(\R')$ can be computed in $O\left(N \cdot \OUT^{\phi_{e_3, e_1}}\right)$ time following Lemma~\ref{lem:base}. To compute $\Q_{e_1,e_2}(\R')$, the Yannakakis algorithm needs to materialize the following intermediate join result:
    $\displaystyle{R_{e_1} \Join \left(\Join_{e_3 \in \mathcal{N}_{e_1} - \{e_2\}} \Q_{e_3, e_1}(\R')\right)}$.
    As there are $N$ tuples in $R_{e_1}$, and each tuple in $R_{e_1}$ can be joined with at most 
    $\displaystyle{\prod_{e_3 \in \mathcal{N}_{e_1} - \{e_2\}} \tOUT^{\phi_{e_3, e_1}}} = \tOUT^{\phi_{e_1, e_2}}$
    tuples in this intermediate result, its size can be bounded by $O\left(N \cdot \OUT^{\phi_{e_1,e_2}}\right)$. The cost of line 14 is bounded by Lemma~\ref{lem:base}. The cost of lines 15-18 is bounded by $O(N)$. All other lines take $O(1)$ time. As analyzed above, the number of the while-loop iterations is $O(2^{|\E|})$, which is still a constant. So, the partitioning procedure takes $O\left(N \cdot \OUT^{1-\frac{1}{\fnfhtw}}\right)$ time. Putting everything together, we obtain Lemma~\ref{lem:partition}.

    \section{Output Size Estimation}
    \label{appendix:output}

    \subsection{Line Queries~\cite{cohen1994estimating} }

    It has been shown how to obtain a constant-factor approximation of $\OUT$ for line queries in near-linear time~\cite{cohen1994estimating}. We borrow the technique of \emph{k minimum values} (KMV) \cite{Beyer:sigmod2007,bar2002counting}, which is more commonly used to estimate the number of distinct elements in the streaming model. KMV works by applying a hash function to the input items, and keeping the $k$ minimum hash values, denoted as $v_1,v_2,\cdots,v_k$.  It has been shown that, with $k = O(\frac{1}{\epsilon^2})$, the estimator $\frac{k-1}{v_k}$ is an $(1+\epsilon)$-approximation of the number of distinct items in the data stream, with at least constant probability. Moreover, given the KMVs of two sets, the KMV of the union of the two sets can be computed by simply ``merging'' the two KMVs, i.e., keeping the $k$ minimum values of the $2k$ values from the two KMVs, provided that they use the same hash function. 

    On a line query, for each $a\in \dom(A_1)$, we will obtain a constant-factor approximation of $\OUT_a = |\pi_{A_{n+1}} R_1(a, A_2) \Join R_2(A_2,A_3) \Join \cdots \Join R_n(A_n, A_{n+1})|$.  Note that $\OUT = \sum_{a\in \dom(A_1)} \OUT_a$.  We compute a hash value for each distinct value in $\dom(A_{n+1})$ to do so. Then it suffices to compute, for each $a\in\dom(A_1)$, a KMV (with a constant $k$) over all distinct values in $\dom(A_{n+1})$ that can join with $a$.  This can be done by 
    using the merge operation above to compute the min.  More precisely, for $i=n, n-1, \dots, 1$, we compute the KMV for each $a\in \dom(A_i)$, by merging all the KMVs for $b \in \dom(A_{i+1})$ such that $(a,b) \in R_i(A_i, A_{i+1})$.

    The KMV obtained from each $a\in \dom(A_1)$ gives us a constant-factor approximation of $\OUT_a$ only with constant probability.  To boost the success probability, we run $O(\log N)$ instances of this algorithm in parallel using independent random hash functions and return the median estimator for each $\OUT_a$.  This boosts the success probability to $1-1/N^{O(1)}$ for each $\OUT_a$.  By a union bound, the probability that all estimators are constant-factor approximations is also $1-1/N^{O(1)}$.  Then, we also have a constant-factor approximation of $\OUT$.  The runtime of this algorithm is $\widetilde{O}(N)$.

    \subsection{Estimate $\OUT$ on the fly \cite{hu2024fast}}

    However, it is unknown how to extend the KMV-based techniques to general acyclic queries. In \cite{hu2024fast}, Hu showed a method that can obtain a $2$-approximation of $\OUT$ on the fly. As shown in Algorithm~\ref{alg:hybrid-yannakakis}, it iteratively compresses the input instances and invokes Algorithm~\ref{alg:partition} as a primitive, which has been described in Section~\ref{sec:acyclic}. 

    \subsection{Double Guess of $\OUT$ \cite{deep2024output}}
    Another simple trick can be applied to estimate the output size of join-aggregate queries~\cite{deep2024output}.  We can doubly guess the value of $\OUT$ starting from $1,2,4,\cdots$. For the $i$-th trial, we run the algorithm for $c \cdot N \cdot (2^i)^{1-\frac{1}{k}}$ time before terminating the simulation if the execution is not completed, for some sufficiently large constant $c$. The overall time complexity is $\sum_{i \in [\lceil \log \OUT \rceil]} c \cdot \left(N \cdot (2^i)^{1-\frac{1}{\fnfhtw(\Q)}} + 2^i\right) = O\left(N \cdot \OUT^{1-\frac{1}{\fnfhtw(\Q)}} + \OUT\right)$.
    
    \begin{algorithm}[h]
    \caption{\textsc{HybridYannakakisWithoutOUT}$(\Q = (\V,\E,\y), \R)$}
    \label{alg:hybrid-yannakakis}
    \SetKwInOut{Input}{Input}
    \SetKwInOut{Output}{Output}
    \Input{A separated acyclic join-aggregate query $\Q$ and an input instance $\R$;}
    \Output{Query result $\Q(\R)$;}
    \lIf{$|\E|=1$, say $\E = \{e\}$}{
    \Return $\oplus_{\V - \y} R_e$}
    $\E' \gets \{e \in \E: e \cap \y \neq \emptyset\}$\;
        $e \gets$ an arbitrary relation in $\in \E'$\;
        Put an ordering on elements in $\pi_{e \cap \y} R_e$ as $a_1,a_2,a_3,\cdots$\;
        $S^{(0)}_e \gets R_e$, $i\gets 1$\;
        \While{\textup{\textbf{true}}}{
        $S^{(i)}_e \gets \emptyset$\;
        \ForEach{$t \in S^{(i-1)}_{e}$}{
        Suppose $\pi_{e \cap \y}t = a_j$\;
        $t' \gets $ a tuple with $\pi_{e \cap \y} t' = a_{\lfloor\frac{j+1}{2}\rfloor}$ and $\pi_{A} t' = \pi_A t$ for any attribute $A \in e - \y$\; 
        $S^{(i)}_{e} \gets S^{(i)}_{e} \cup \left\{t'\right\}$\;
        }
        \lIf{$\left|\pi_{e \cap \y} S^{(i)}_e\right|=1$}{\textbf{break}}
        $i\gets i+1$\;
    }
    $R_{e'} \gets R_{e'} \ltimes S^{(i)}_{e}$\;
    $\Q' \gets \left(\V - e \cap \y, \E - \{e\}, \y - e\right)$\;
    $\mathcal{J}^{(i)} \gets \left(\pi_{e \cap \y}S^{(i)}_e\right) \times \textsc{HybridYannakakisWithoutOUT}\left(\Q',\R - \{R_e\}\right)$\;
    \While{$i > 0$}{
         $\mathcal{J}^{(i-1)} \gets \textsc{HybridYannakakis}\left(\Q, \R - \{R_e\} + \left\{S^{(i-1)}_{e}\right\}, 2 \cdot |\mathcal{J}^{(i)}|\right)$\; \text{// Algorithm~\ref{alg:partition}}
         $i\gets i-1$\;
    }
    \Return $\mathcal{J}^{(0)}$\; 
    \end{algorithm}

\end{document}